\documentclass[12pt]{article}

\usepackage{scicite}
\usepackage{times}
\usepackage{caption}
\usepackage{graphicx} 
\usepackage{multirow}
\usepackage{newfloat}
\usepackage{ragged2e}
\usepackage[version=3]{mhchem} 
\DeclareCaptionLabelFormat{Sfignum}{#1~S#2}
\DeclareFloatingEnvironment[fileext=lof,listname={List of fragments}, 
                            name=Movie, placement=h]{movie}
\newenvironment{sciabstract}{%
\begin{quote} \bf}
{\end{quote}}

\title{High-resolution spatio-temporal strain imaging reveals loss mechanisms in a surface acoustic wave device} 

\author
{Tao Zhou,$^{1,2}$ Alexandre Reinhardt,$^{3,\ast}$ Marie Bousquet,$^{3}$ Joel Eymery,$^{4}$ \\ Steven Leake$^{1}$, Martin V. Holt,$^{2}$ Paul G. Evans,$^{5,\ast}$ Tobias Schülli,$^{1,\ast}$\\
\\
\normalsize{$^{1}$ESRF - The European Synchrotron, Grenoble, 38000, France}\\
\normalsize{$^{2}$Center for Nanoscale Materials, Argonne National Laboratory, Lemont, IL 60439, USA}\\
\normalsize{$^{3}$Université Grenoble Alpes, CEA-LETI, Grenoble, 38000, France}\\
\normalsize{$^{4}$Université Grenoble Alpes, CEA-IRIG-MEM-NRX, Grenoble, 38000, France}\\
\normalsize{$^{5}$University of Wisconsin-Madison, Madison, WI 53706, USA}\\
\\
\normalsize{$^\ast$To whom correspondence should be addressed.} \\
\normalsize{E-mail: alexandre.reinhardt@cea.fr, pgevans@wisc.edu, schulli@esrf.fr}
}

\date{}

\begin{document}

The submitted manuscript has been created by UChicago Argonne, LLC, Operator of Argonne National Laboratory(“Argonne”). Argonne, a U.S. Department of Energy Office of Science laboratory, is operated under Contract No. DE-AC02-06CH11357. The U.S. Government retains for itself, and others acting on its behalf, a paid-up nonexclusive, irrevocable worldwide license in said article to reproduce, prepare derivative works, distribute copies to the public, and perform publicly and display publicly, by or on behalf of the Government.
\newpage


\baselineskip24pt


\maketitle 


\begin{sciabstract} 
% 150 words limit, currently 144
Surface acoustic wave devices are key components for processing radio frequency signals in wireless communication because these devices offer simultaneously high performance, compact size and low cost. The optimization of the device structure requires a quantitative understanding of energy conversion and loss mechanisms. Stroboscopic full-field diffraction x-ray microscopy studies of a prototypical one-port resonator device revealed the existence of unanticipated acoustic loss. A non-uniform acoustic excitation in the active area was responsible for the substantial end and side leakages observed at the design frequency. Quantitative analysis of the strain amplitude using a wave decomposition method allowed the determination of several key device parameters. This high-resolution spatiotemporal strain imaging technique is, more generally, suited for studying nanophononics, specifically when the feature size is smaller than optical wavelengths. The strain sensitivity allows precise measurement of acoustic waves with picometer-scale amplitude.

\end{sciabstract}
\section*{Introduction}
Surface acoustic wave (SAW) devices underpin radio frequency electronics applications such as signal filtering because these devices can be simultaneously compact and inexpensive\cite{Campbell1989}. There is also significant interest in the use of SAW devices as highly sensitive gas\cite{Kumar2022} or biosensors \cite{Lange2008}, with the possibility of direct integration into lab-on-a-chip platforms\cite{Yeo2014}. SAW devices also promise a large variety of intriguing nanoscale applications ranging from straintronics \cite{Delsing2019} to quantum communication\cite{Dumur2021, Andersson2022}. High-spatial-resolution characterization is critical for both application-driven and fundamental research in all of these applications. The transduction from electric energy to mechanical energy and vice versa in SAW devices is typically achieved with interdigital transducers (IDTs)\cite{White1965, Manenti2017}, which are two interlocking comb-shaped arrays of metallic electrodes on a crystalline piezoelectric substrate. One of the transducer electrodes is grounded while an oscillating voltage is applied to the other, resulting in a spatially and temporally periodic strain field near the surface of the piezoelectric crystal. The strain field propagates at speed in the range of a few km/s, forming an acoustic wave at the surface. The oscillation frequency of the SAW is in the 100 - 1000 MHz range, and is typically too fast for even state-of-the-art high speed scientific cameras\cite{Manin2018}.

The mechanisms and magnitude of acoustic loss are critical in determining the performance of SAW devices\cite{Koskela1999}. Loss mechanisms are studied with electrical measurements by using the frequency response of the electrical parameters, most commonly the scattering $S$, impedance $Z$ and admittance $Y$-parameters\cite{Shu2016, Kittmann2018}. While highly sensitive to loss, those electrical parameters provide information only at the scale of the entire device and lack the spatial or temporal resolution necessary to understand the origins of the loss. Spatially resolved information can be acquired using scanning probe techniques with optical \cite{Knuuttila2000, Hisatomi2023}, mechanical \cite{Hesjedal1998} or x-ray \cite{Hanke2023} probes. The spatial resolution of optical measurements is ultimately limited to hundreds of nanometers by the optical diffraction limit. Atomic resolution is in theory achievable with mechanical measurements, but the applicable frequency range is severely limited by the cantilever resonance to below 1 MHz. Response in the time domain can be further obtained by synchronizing the probe to the SAW excitation, in a scheme known as stroboscopic imaging. Stroboscopic optical\cite{Ludvigsen2015} and x-ray\cite{Nicolas2014, Reusch2013} methods have been able to demonstrate a time resolution on the order of 100 ps, sufficient for studying SAW devices of as fast as 5 GHz at their Nyquist frequency. Stroboscopic scanning diffraction x-ray microscopy can have a spatial resolution as small as 25 nm, with a relatively small field of view (FoV) due to time spent on raster scans \cite{Whiteley2019}. These state-of-the-art methods have shown a sensitivity to the surface displacement on the order of 100 pm\cite{Knuuttila2000, Hesjedal1998, Hanke2023, Ludvigsen2015, Nicolas2014, Reusch2013, Whiteley2019}. X-ray diffraction methods have the potential for significantly higher sensitivity because they measure directly the acoustically induced strain waves. However, such higher sensitivity has not been realized due to the absence of necessary analytical techniques or data reduction methods.
 
We report the use of stroboscopic full field diffraction x-ray microscopy (s-FFDXM) and an associated wave decomposition analysis method that, together, achieve high-resolution spatiotemporal imaging of the acoustically induced strain waves. The method has a strain sensitivity of $\sim10^{-7}$, corresponding to a surface displacement of 1 pm, that is a factor of 100 times better than most existing methods. The time resolution is 100 ps. Measurements on a prototypical one-port SAW resonator reveal multiple mechanisms of acoustic loss with distinct spatial and time dependencies. The relative strength of these loss mechanisms as well as several key device parameters were determined with quantitative analysis. The $450\times250~\mu$m$^2$ FoV of s-FFDXM is comparable to those of optical full-field methods\cite{Lipiainen2015, Telschow2003}, which enables concurrent characterizations of multiple device regions exhibiting diverse spatio-temporal behavior.

\section*{Stroboscopic Strain Imaging of Surface Acoustic Waves}
\begin{figure}[hbt]
\centering
\includegraphics[width=1\textwidth]{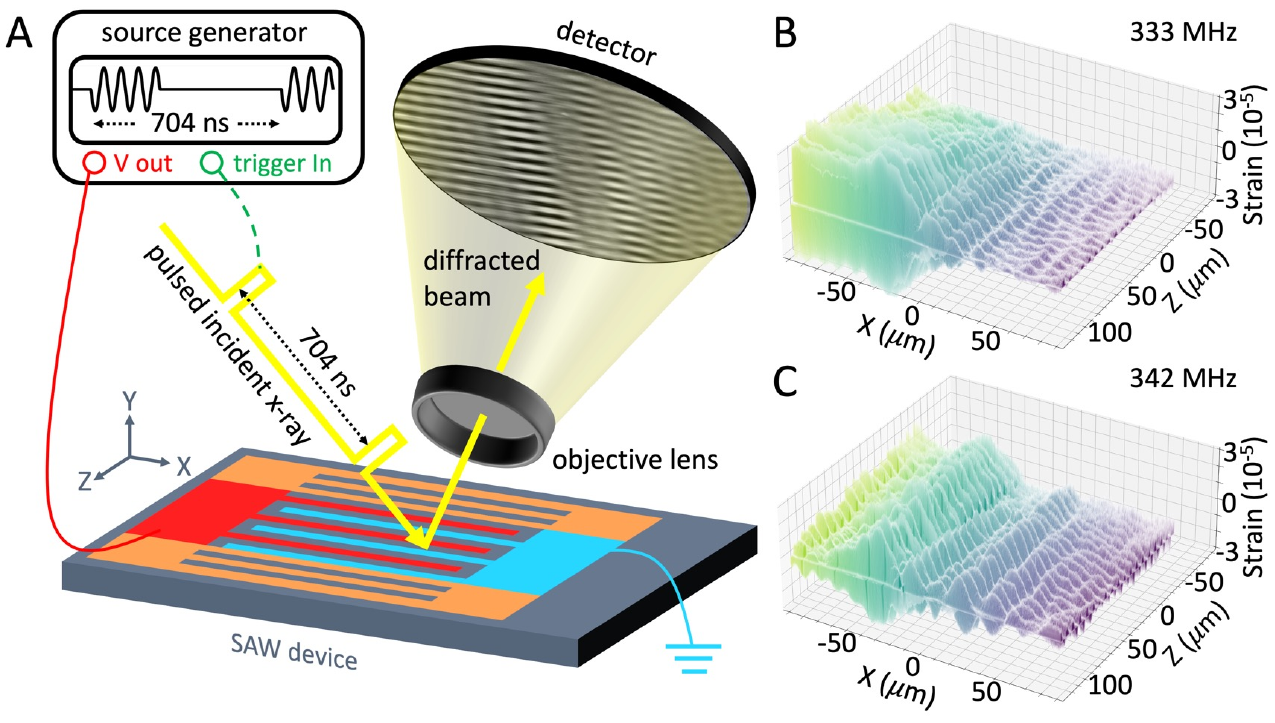}
\caption{\label{fig:sch} \textbf{Stroboscopic full-field diffraction x-ray microscopy.} (A) The SAW excitation was electronically synchronized to the x-ray pulses with a tunable delay. A dark-field image was formed on a two-dimensional x-ray detector by projecting a magnified image of the diffracted beam using an x-ray objective lens. Analysis of the dark field images acquired during a $\theta$-$2\theta$ scan allowed visualization of the strain wave, as shown in (B) and (C). The reflector gratings and the two IDT electrodes appear in orange, red, and blue, respectively. (B) Instantaneous strain map for an excitation at 333 MHz showing a uniform standing wave in the resonator area ($X<0$, $Z>0$), one frame from Movie S1. (C) Instantaneous strain map for an excitation at 342 MHz showing strong leakage into the reflector ($Z<0$) and into the bus bar ($X>0$), one frame from Movie S2..}
\end{figure}

Figure \ref{fig:sch}A shows a schematic of the s-FFDXM experiment. A single-port synchronous SAW resonator device\cite{Koskela2000} was fabricated on a Y-cut \ce{LiNbO3} substrate with a design wavelength of $\lambda_\text{SAW} = 10~\mu$m and a design resonance frequency of 339 MHz. Applying an oscillating electric field between adjacent IDT fingers generated a series of SAWs propagating along the \textbf{+Z} and \textbf{--Z} crystalline direction. These SAWs were subsequently reflected by the gratings at the ends of the IDT array, leading to the formation of a standing wave in the resonator area. 

The excitation for the stroboscopic imaging experiments was synchronized to the x-ray pulses. A burst consisting of 64 periods of a sine waveform were generated with a repetition rate of 1.42 MHz (704 ns) and a peak-to-peak voltage of 3 V. The time-dependent response of the SAW device was studied by shifting the start time of the electrical burst with regard to the x-ray pulses using an electronic delay. At a fixed delay $t$, x-ray photons from different pulses probed the same time dependent event at time $t$ after the start of the electrical signal. The temporal resolution was ultimately limited by the x-ray pulse width to 100 ps FWHM\cite{timeres}.

The SAW device was illuminated with a quasi-parallel x-ray beam with an incident convergence angle of $10^{-5}$ rad. The pseudo-cubic 300 reflection of the \ce{LiNbO3} substrate was chosen for the diffraction experiments. The intensity and wavevector of this reflection carry information about the atomic displacements along the surface normal (\textbf{Y}) direction. The diffracted beam was collected by an objective lens to form a dark field image on the detector. The effective pixel size was 114 nm. A large FoV was imaged with each detector acquisition, which allowed simultaneous strain imaging of multiple device areas at various excitation frequencies (Figure \ref{fig:sch}B and C).

\section*{Results}
\paragraph*{s-FFDXM Observation of the Resonance\\}
\begin{figure}[ht]
\centering
\includegraphics[width=1\textwidth]{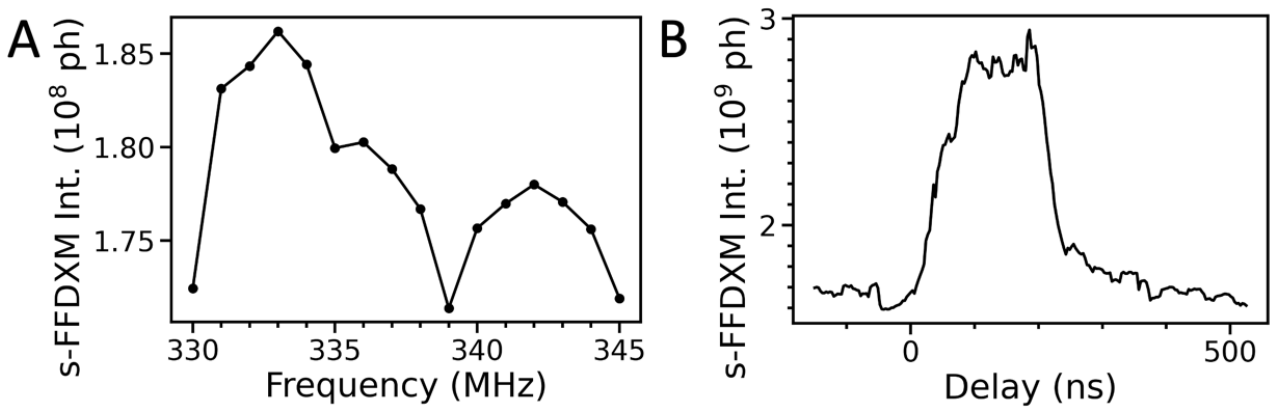}
\caption{\label{fig:rso} \textbf{s-FFDXM observation of SAW resonance.} (A) Integrated s-FFDXM intensity from the resonator as a function of SAW excitation frequency. The baseline with excitation was $1.7\times10^8$ photons measured at an excitation frequency of 350 MHz. The baseline without excitation was $1.1\times10^8$ photons. (B) Integrated s-FFDXM intensity as a function of the electronic delay $t$. The SAW was excited with a burst of 64 periods of a sine waveform at 333 MHz. The start of the burst is at $t=0$.}
\end{figure}

The device performance was first evaluated using the integrated s-FFDXM intensity. As explained in the \textbf{Methods}, the integrated intensity is proportional to the magnitude of the local curvature of the piezoelectric substrate, which is in turn proportional to the SAW amplitude\cite{Kalman1983}. Figure \ref{fig:rso}A shows the integrated s-FFDXM intensity from the resonator region as a function of SAW excitation frequency. A dip in the intensity at 339 MHz indicates on average a weaker SAW amplitude, which is expected as 339 MHz corresponds to the resonance frequency at which the device exhibits minimum impedance (maximum admittance in Figure~S\ref{fig:sup_imp}). A peak in the intensity at 342 MHz indicates a higher SAW amplitude, which is also expected because 342 MHz is the anti-resonance frequency at which the device exhibits maximum impedance (minimum admittance in Figure~S\ref{fig:sup_imp}). However, Figure \ref{fig:rso}A indicates that the maximum SAW amplitude was instead observed at an off-resonance frequency of 333 MHz. A spatiotemporal analysis of the acoustic response of the device, below, reveals the origin of the s-FFDXM intensity maxima.

\paragraph*{Two-dimensional Wave Pattern in the Resonator\\}
\begin{figure}[ht]
\centering
\includegraphics[width=0.5\textwidth]{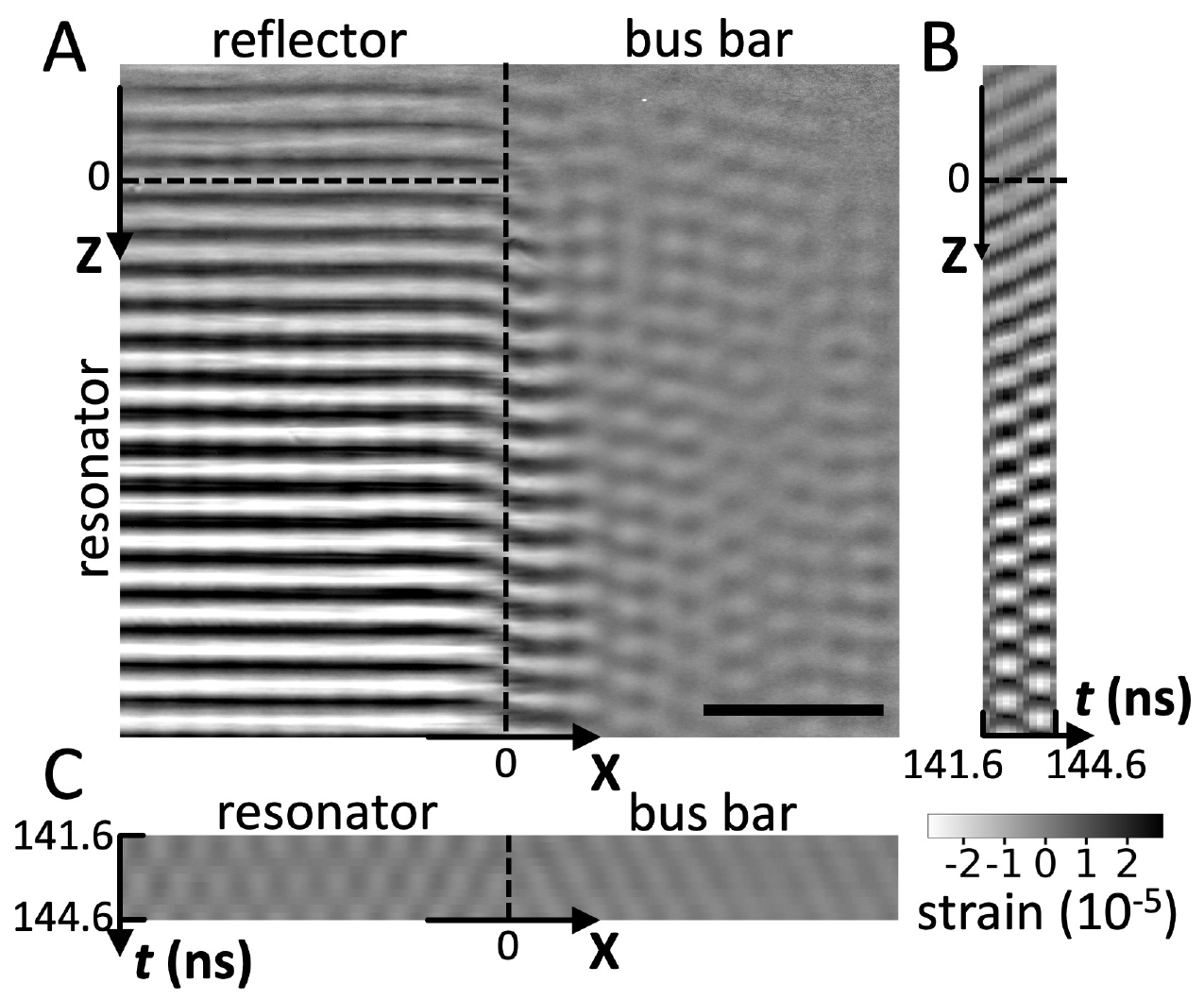}
\caption{\label{fig:333} \textbf{Strain imaging of acoustic response at 333 MHz.} (A) Strain map of the SAW device at $t = 144$ ns. The resonator is at the lower left corner of the imaged area, in the region of positive $Z$ and negative $X$. The top reflector occupies the region of negative $Z$ and negative $X$. The bus bar is in the region with $X>0$. The scale bar is 50 $\mu$m. (B) Primary excitation in the region near $X=-100~\mu$m from $t =$ 141.6 to 144.6 ns. The spatiotemporal strain map in (B) spans the same range in $Z$ as the map in (A). (C) Parasitic excitation in the region near $Z=30~\mu$m from $t =$ 141.6 to 144.6 ns. The spatiotemporal strain map shares the same $X$ coordinates as (A). }
\end{figure}

The acoustic response of the SAW device was studied in detail at 333 MHz. As shown in Figure \ref{fig:rso}B, the s-FFDXM intensity increased at the start of the burst at $t=$ 0 ns and reached a maximum after 25 periods. The intensities then remained constant before decreasing after the end of the burst at $t=$ 192 ns. Figure~\ref{fig:333}A shows the instantaneous strain map of the device at $t=$ 144 ns, corresponding to the beginning of the 48$^\text{th}$ burst period. The color scale is centered at zero strain, such that brighter and darker shading indicate compressive and tensile strain, respectively. The maximum amplitude was $3\times10^{-5}$. The strain values were determined using the center-of-mass analysis for 31 s-FFDXM images acquired during a $\theta$-$2\theta$ x-ray scan. Details of the analysis are further described in the \textbf{Methods} section. 

The strain map in Figure~\ref{fig:333}A reveals a two-dimensional wave pattern in the resonator area. Figure~S\ref{fig:sup_fft} shows the Fourier transform of the two-dimensional wave pattern. The strain amplitude had a spatial period of $10.0\pm0.06~\mu$m along the \textbf{Z} direction. We refer to this modulation as the primary excitation because both the direction of the modulation and its spatial period match the designed acoustic response of the device. A modulation of the strain amplitude along the \textbf{X} direction was not intended in the design and is referred to as the parasitic excitation. At $11.06\pm0.11~\mu$m, the spatial period of the parasitic excitation was 10.6\% larger than what was observed for the primary excitation, the cause of which is further explored below. 

\paragraph*{End Leakage of the Primary Excitation\\}
Figure~\ref{fig:333}B shows the spatiotemporal strain evolution of the primary excitation wave. Information in the time domain was acquired by combining strain maps (given in Movie S3) that were measured at 11 evenly spaced delay values spanning the 3 ns period of the 48$^\text{th}$ acoustic burst excitation. The contribution of the parasitic excitation was suppressed in Figure~\ref{fig:333}B by averaging the strain over one spatial period along the \textbf{X} direction. A propagating wave was observed in the reflector area at $Z<0$. With increasing $t$, the peaks and valleys of the strain wave were shifted to smaller $Z$ values, indicating that the observed propagating wave was in fact end leakage into the reflector area traveling in the \textbf{--Z} direction. 

Away from the reflector, at $Z>50~\mu$m, a standing wave pattern was observed in the resonator area. The positions of the IDT electrodes and reflector gratings were determined using methods described in Figure~S\ref{fig:sup_ecd}, and are schematically shown in Figure~S\ref{fig:sup_333v}A. The nodes (i.e., with minimum strain amplitude) of the standing wave were centered on the IDT electrodes, while the anti-nodes (i.e., with maximum strain amplitude) were located equidistant from two adjacent IDT electrodes. The position of the nodes and anti-nodes confirms that the primary excitation was the designed acoustic response of the device, which operated at the entrance of the stop-band for the Rayleigh wave in the electrodes grating. 

Wave decomposition analysis was performed on the primary excitation by fitting the strain amplitude of a wave model to the experimentally observed spatiotemporal strain evolution. Procedures for the fitting as well as descriptions of the wave model are listed in the \textbf{Methods} section. Figure~S\ref{fig:sup_333v}B shows the best fit result of the wave model, in excellent agreement with the experimental data relayed in Figure~S\ref{fig:sup_333v}C. At $Z < 100~\mu$m, the spatiotemporal strain evolution can be simply described by the superposition of two waves traveling in opposite directions. Figures~S\ref{fig:sup_333v}D and E show, respectively, the amplitude of $\psi_{z+}$, the wave traveling in the \textbf{+Z} direction and the amplitude of $\psi_{z-}$, the wave traveling in the \textbf{--Z} direction. In the resonator area ($Z>0$), the maximum amplitude of $\psi_{z-}$ was 40\% stronger than for $\psi_{z+}$. In the top reflector area ($Z<0$), the amplitude of the reflected wave $\psi_{z+}$ was negligible, which indicated a low reflectivity of the grating at 333 MHz. Similarly, the amplitude of the transmitted wave $\psi_{z-}$ was high into the reflector area, signifying a substantial end leakage.

The model consisting of only two traveling waves did not explain the spatiotemporal strain evolution in the lower part of the resonator at $Z > 100~\mu$m. Matching the superposition of $\psi_{z-}$ and $\psi_{z+}$ with the spatiotemporal strain evolution in Figure~\ref{fig:333}B revealed the presence of a third wave component $\psi_{t}$. As shown in Figure~S\ref{fig:sup_333v}F, $\psi_{t}$ oscillated in time at twice the frequency of the electrical burst excitation. The maximum amplitude of $\psi_{t}$ was observed when the amplitude of applied electric field was maximized, regardless of the voltage polarity. Additionally, the amplitude of $\psi_{t}$ did not exhibit a periodic oscillation in space, indicating that it was not caused by higher harmonics generation. Further evidence supporting the existence of $\psi_{t}$ can be found in Figure~S\ref{fig:sup_fit}.

\paragraph*{Origin of the Parasitic Excitation\\}
Figure~\ref{fig:333}C shows the spatiotemporal strain evolution of the parasitic excitation wave. The contribution of the primary excitation to the map in Figure~\ref{fig:333}C was suppressed by averaging the strain over one spatial period along the \textbf{Z} direction. The strain amplitude of the parasitic wave was a factor of ten weaker than the primary wave, which explained the lower strain amplitude observed in Figure~\ref{fig:333}C. The spatiotemporal strain evolution of the parasitic excitation wave can be described by the superposition of two waves, $\psi_{x+}$ travelling in the \textbf{+X} direction and $\psi_{x-}$ travelling in the \textbf{--X} direction. Their summed amplitude, shown in Figure~S\ref{fig:sup_333h}A, is an excellent match for the experimental data relayed in Figure~S\ref{fig:sup_333h}B. A standing wave pattern was observed in the resonator area for $X<0$, while a pure propagating wave was observed in the bus bar area for $X>0$. With increasing $t$, the peaks and valleys of the propagating wave were shifted to larger $X$ values. This indicates that the observed propagating wave was in fact side leakage into the bus bar area traveling in the \textbf{+X} direction.

The wave decomposition in Figure~S\ref{fig:sup_333h}C and D shows that at any given time $t$, $\psi_{x+}$ and $\psi_{x-}$ were in phase at $X=0$, leading to the conclusion that the source of the parasitic excitation was at the boundary separating the resonator and the bus bar. The maximum strain amplitude of both $\psi_{x-}$ and $\psi_{x+}$ were found at $X=0$, which further supports this explanation. At $X>0$, the strain amplitude of $\psi_{x+}$ decreased as the wave propagated further away from its source of origin. The part of $\psi_{x+}$ at $X<0$ was generated at the other boundary  located on the other side of the resonator at $X = -300~\mu$m, its strain amplitude also decreasing as the wave propagated further away from its source of origin. We hypothesize that the parasitic excitation was generated by the electric field between the tip of one IDT finger and the opposite busbar, as illustrated in Figure~S\ref{fig:sup_ori}. The generation of the parasitic wave via this mechanism explains the much weaker amplitude of the parasitic excitation compared to the primary excitation because there were much fewer electrodes for generating $\psi_{x\pm}$ than for generating $\psi_{z\pm}$. 

Having established the direction in which the parasitic excitation wave propagates, we can now explain the 10.6\% larger modulation period compared to the primary excitation. The period of the modulation is proportional to the velocity of the SAW, which varies along different crystallographic directions due to anisotropic elasticity\cite{Hossain2019}. The ratio of observed modulation period is consistent with previous pulse inference measurements which indicated that the SAW velocity along the \textbf{X} direction is 12\% faster than along the \textbf{Z} direction\cite{Kushibiki1999}.

\paragraph*{Acoustic Response at the Anti-Resonance Frequency\\}
\begin{figure}[ht]
\centering
\includegraphics[width=0.5\textwidth]{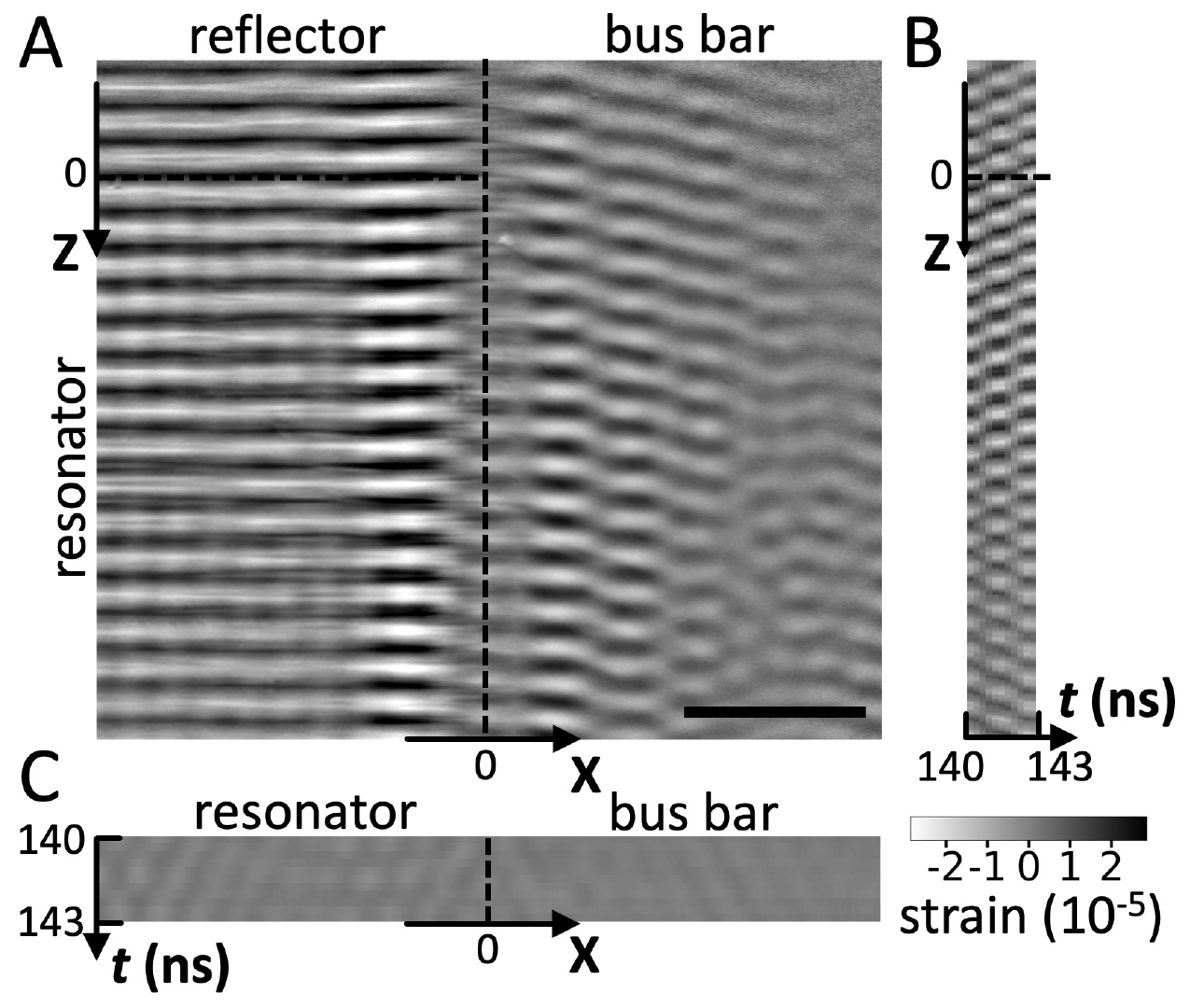}
\caption{\label{fig:342} \textbf{Strain imaging of acoustic response at 342 MHz.} (A) Strain map of the SAW device at $t = 140$ ns. The resonator is at the lower left corner of the imaged area, in the region of positive $Z$ and negative $X$. The top reflector occupies the region of negative $Z$ and negative $X$. The bus bar is at $X>0$. The scale bar is 50 $\mu$m. (B) Primary excitation in the region near $X=-100~\mu$m from $t =$ 140 to 143 ns.  The spatiotemporal strain map in (B) spans the same range in $Z$ as the map in (A). (C) Parasitic excitation in the region near $Z=30~\mu$m from $t =$ 140 to 143 ns. The spatiotemporal strain map in (C) spans the same range in $X$ as the map in (A).}
\end{figure}

The acoustic response of the SAW device was also studied at the anti-resonance frequency of 342 MHz. Figure~\ref{fig:342}A shows an instantaneous strain map of the device at $t=$ 140 ns, corresponding to the beginning of the 48$^{\text{th}}$ burst period at 342 MHz. The strain map at the anti-resonance differs from the previous map at 333 MHz in several ways. First, a less uniform distribution of the SAW amplitude was observed in the resonator area at 342 MHz compared to 333 MHz. The SAW amplitude was much stronger in the narrow area within 40 $\mu$m from the bus bar than in the rest of the resonator. Second, standing waves were observed in the reflector area, as evident from the spatiotemporal strain evolution shown in Figure~\ref{fig:342}B at $Z<0$. The presence of standing waves indicates more propagating waves were reflected back by the grating, which is indicative of a high reflectivity of the reflector grating at 342 MHz. Third, and most importantly, Figure~\ref{fig:342}A shows a much stronger wave propagating into the bus bar area. The leakage into the bus bar area consists of mainly the primary excitation, which are waves with a spatial periodicity along the \textbf{Z} direction. The amplitude of the parasitic excitation (i.e., waves with a spatial periodicity along the \textbf{X} direction) was actually weaker than at 333 MHz, as is evident from the spatiotemporal strain map shown in Figure~\ref{fig:342}C.

\paragraph*{Quantitative comparison of the wave excitation\\}
\begin{figure}[ht]
\includegraphics[width=1\textwidth]{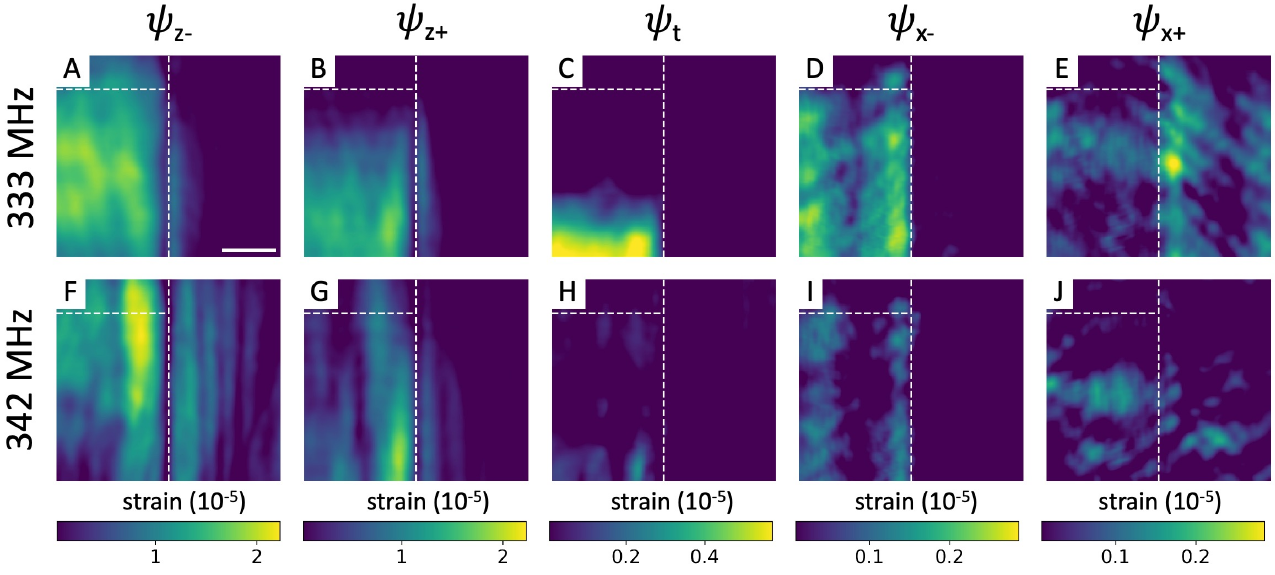}
\caption{\label{fig:comparison} \textbf{Amplitude map of the wave components at 333 and 342 MHz}. Strain amplitude map at 333 MHz for (A) $\psi_{z-}$, (B) $\psi_{z+}$, (C) $\psi_t$, (D) $\psi_{x-}$ and (E) $\psi_{x+}$. Strain amplitude map at 342 MHz for (F) $\psi_{z-}$, (G) $\psi_{z+}$, (H) $\psi_t$, (I) $\psi_{x-}$ and (J) $\psi_{x+}$. The color scale presented in each column applies to both 333 MHz and 342 MHz. The imaged areas are the same as in Figure~\ref{fig:333}a and Figure~\ref{fig:342}a. The boundaries separating the resonator, reflector and bus bar are marked with dashed lines. The scale bar is 50 $\mu$m.}
\end{figure}

The results of the wave decomposition analysis for excitation frequencies of 333 and 342 MHz are shown in Figure~\ref{fig:comparison}. The spatially resolved strain amplitudes, extracted individually for each wave component, allowed a quantitative comparison of the relative strength of the side and end leakages, as well as the determination of key device parameters such as the standing wave ratio (SWR) and the grating reflectivity.

At 333 MHz, a standing wave with uniformly distributed amplitude was formed by the two primary wave components $\psi_{z-}$ (Figure~\ref{fig:comparison}A) and $\psi_{z+}$ (Figure~\ref{fig:comparison}B) in the resonator area. An amplitude mismatch was observed between $\psi_{z-}$ and $\psi_{z+}$, corresponding to a SWR of 6:1. The reflectivity of the top grating was 12\%, calculated using the amplitude ratio in the reflector area between the transmitted wave $\psi_{z-}$ and reflected wave $\psi_{z+}$. A low reflectivity was expected at 333 MHz due to the synchronous design of the resonators, where the same periodicity was used for both the reflector and transducer electrodes. The experimentally determined reflectivity was in good agreement with our finite element modeling which predicted a reflectivity of 10\% at 333 MHz. We further note that the grating reflectivity cannot be directly measured with electrical methods. The maximum amplitude of $\psi_t$ was a factor of 3 times weaker than $\psi_{z\pm}$. From Figure~\ref{fig:comparison}C, we confirm that $\psi_t$ was only found at the bottom part of the resonator. Although the maximum wave amplitude of $\psi_{x-}$ (Figure~\ref{fig:comparison}D) and $\psi_{x+}$ (Figure~\ref{fig:comparison}E) were a factor of 2 smaller than $\psi_t$, the parasitic excitations $\psi_{x\pm}$ were apparently responsible for higher loss compared to $\psi_t$ due to their presence in a larger area of the device.

At 342 MHz, the maximum wave amplitude of the primary excitations was 11\% higher than at 333 MHz. However, SAW with strong wave amplitude was only observed in a narrowly confined area within 40 $\mu$m from the boundary separating the resonator and the bus bar. As a result, the area averaged SAW amplitude in the resonator was actually 20\% weaker at 342 MHz than at 333 MHz, consistent with the observed s-FFDXM intensity difference in Figure~\ref{fig:rso}A. The calculated SWR of 6:1 at 342 MHz was the same as at 333 MHz. The persistently low SWR value indicated that a true standing wave was not achievable in the resonator area under continuous excitations. A higher SAW amplitude was observed in the reflector at 342 MHz. The maximum strain amplitude of the end leakage $\psi_{z-}$ (Figure~\ref{fig:comparison}F, reflector area) was a factor of 3 higher at 342 MHz than at 333 MHz. The substantial end leakage observed at 342 MHz was despite a high reflectivity of the grating, measured at 29\% using the amplitude of the reflected wave $\psi_{z+}$ (Figure~\ref{fig:comparison}G, reflector area). A substantial amount of side leakage was also observed. The leaked primary excitation $\psi_{z-}$ propagated in the \textbf{+X} direction into the bus bar, to a distance more than 5 times further at 342 MHz than at 333 MHz. The wave amplitude for $\psi_t$ (Figure~\ref{fig:comparison}H) and for the parasitic excitations $\psi_{x-}$ (Figure~\ref{fig:comparison}I) and $\psi_{x+}$ (Figure~\ref{fig:comparison}J) were actually weaker at 342 MHz than at 333 MHz, which is understandable as the device was designed to operate close to 342 MHz.

\section*{Discussion}
In this work, we have studied operando the loss mechanisms in a single-port surface acoustic wave resonator device using stroboscopic full field diffraction x-ray microscopy. The integrated s-FFDXM intensities were proportional to the SAW amplitude, which indicated the presence of a stronger SAW in the resonator at an off-resonance frequency of 333 MHz instead of at the expected frequency of 342 MHz. Detailed spatiotemporal analysis showed that much of the SAW excited at 342 MHz was narrowly confined within 40 $\mu$m from the edge of the resonator. The non-uniformity of the excited SAW led to substantial side and end leakage. For comparison, a standing wave with uniformly distributed amplitude was excited at 333 MHz. The result is the observation of a more efficient electrical-to-mechanical energy conversion at 333 MHz in spite that the IDTs (higher maximum strain amplitude) and the gratings (higher reflectivity) were optimized for 342 MHz. The discrepancy between the electrical and s-FFDXM measurement highlights the importance of temporally and spatially resolved technique for SAW device characterization. Additionally, we have discovered acoustic loss in modes described above by $\psi_t$, $\psi_x-$ and $\psi_x+$ using wave decomposition analysis. The origin of $\psi_t$ is unknown, although the spatiotemporal dependence of $\psi_t$ at 333 MHz could indicate that it was caused by Joule heating. Both $\psi_x-$ and $\psi_x+$ were generated by the electric field between the tip of one IDT finger and the opposite busbar. The strain amplitude of these wave components were relatively small. However, their discovery and quantitative understanding remain crucial to their mitigation through design optimizations.

s-FFDXM in this work has demonstrated a high strain sensitivity. In the \textbf{Methods} section we have estimated the strain sensitivity of our technique to be on the order of $10^{-7}$. The high sensitivity was attributed to a combination of many factors. Matching the penetration depths of the x-ray probe and the excited SAW led to a suppression of the diffraction signal from the unstrained bulk \ce{LiNbO3}. As is shown in Figure~S\ref{fig:sup_radial}, the instrument broadening of the Bragg reflection and low uncertainty due to high photon count rate further allowed precise determination of the peak shifts at far below the angular step size of the measurement. The high strain sensitivity is essential for the observation of weak acoustic losses. For a device with a period of $\lambda_\text{SAW} = 10~\mu$m, the penetration depth of the SAW is also about $10~\mu$m (see Figure~S\ref{fig:sup_pen}). A SAW induced strain of $10^{-7}$ in this case corresponds to a surface displacement of 1 pm (see Figure~S\ref{fig:sup_disp}), which is about 100 times smaller than the detection limit of most x-ray and optical methods\cite{Knuuttila2000, Hesjedal1998, Ludvigsen2015, Nicolas2014, Reusch2013, Whiteley2019}. The wave decomposition analysis developed in this work is critical to the identification of unknown loss mechanisms. Residual analysis of the fitted model was essential to revealing the acoustic loss $\psi_t$ in the resonator area. Conventional loss analysis considered the wave amplitude, and are mostly sensitive to leakage out of the resonator area\cite{Inoue2007, Holmgren2007}. The detection of loss within the resonator area is difficult because the observed wave amplitude is typically dominated by the much stronger primary excitation. Fourier analysis can reveal weak loss overlapping with the strong primary excitation, but only works if the loss oscillates in space at a multiple of the fundamental frequency\cite{Howell2001}. In comparison, the residual analysis used in this work detects loss with any spatial and temporal dependence. 

s-FFDXM complements electrical characterization by offering a high-resolution spatiotemporal strain imaging method for the quantitative discovery of new loss mechanisms. Beyond the system investigated in this work, s-FFDXM is generally applicable to any nanoscale phononic devices, particularly in light of the growing trend towards device miniaturization\cite{Kirchhof2023}. With a wide FoV of 450$\times$250 $\mu$m$^2$, s-FFDXM allows spatiotemporal imaging of multiple device areas at the same time. The time resolution of 100 ps is sufficient for studying devices of up to 5 GHz at their Nyquist frequency. The ultimate time resolution is below 100 fs if performed at a free electron laser facility\cite{Rivas2022}. The spatial resolution of 114 nm is a few times better than what is offered by optical methods, and can be potentially improved to below 10 nm\cite{Bajt2018}. The high spatiotemporal resolution and high strain sensitivity would, for instance, allow operando imaging of Bragg-type phononic crystals with a lattice parameter of 10 nm, corresponding to the manipulation of high-frequency phonons in the 100 GHz regime\cite{PC1}.

\section*{Methods\label{Methods}}
\paragraph*{Surface Acoustic Wave Device Fabrication\\}
Black Y-cut \ce{LiNbO3} wafers were immersed in Caro's acid (\ce{H2SO4}:\ce{H2O2} 9:1) for 10 min and then rinsed in deionized (DI) water at 80°C. The wafers were then immersed in an RCA surface clean 1 solution, \ce{H2O2}:\ce{NH4OH}:\ce{H2O}, for 10 min at 70°C before a second rinsing in DI water at 80°C, followed by a quick dump rinse and drying. A uniform aluminum-silicon (Al$_{0.98}$Si$_{0.02}$) alloy layer with a thickness of 100 nm was deposited using sputter deposition. Bis-trimethylsilyl-amine was used to promote the adhesion of photoresist (UV4, 450 nm) on the Al$_{0.98}$Si$_{0.02}$. After spin coating, the photoresist was baked at 130°C for 120 s. A deep-ultraviolet (UV) mask aligner was used in hard contact mode for pattern exposure. A post-exposure bake was then performed at 130°C for 60 s, before development for 15 s in AZ-326MIF and rinsing in DI water and drying. The Al$_{0.98}$Si$_{0.02}$ was wet-etched in a commercial aluminum etch solution for 195 s, before stripping the photoresist in acetone. The wafers were rinsed in alcohol, then in DI water, and dried.

A single-port SAW resonator design was realized, with one IDT placed between two reflective gratings. The IDT consists in an array of 51 interdigital electrodes aligned perpendicular to the main flat of the Y-cut \ce{LiNbO3} wafer to allow propagation along the \textbf{Z} axis. The labeling of the directions follows the IEEE convention, with \textbf{X}, \textbf{Y}, \textbf{Z} corresponding to respectively the [110], [100] and [001] crystallographic direction \cite{Bartasyte2017}. The crystallographic orientation of the sample was also verified with pole figure measurements. The two gratings, or reflectors, are each made of 50 shorted Al$_{0.98}$Si$_{0.02}$ fingers. All the fingers are exactly 100 nm thick and 2.6 $\mu$m wide, with a pitch size of 5 $\mu$m. Assuming a SAW velocity of approximately 3.4 km/s\cite{Kuppel2002}, the central resonance frequency is expected at about 340 MHz.

The functionality of the resonators was first checked with on-wafer measurements using Ground-Signal-Probes connected to an Agilent N5230A vector network analyzer (VNA). The \ce{LiNbO3} wafer was then diced into small chips containing individual SAW devices, which were subsequently glued on a Printed Circuit Board (PCB) and wire-bonded. The functionality of the SAW devices was checked again with the VNA to ensure that the wire-bonds and the PCB did not significantly affect the electrical response.

\paragraph*{Stroboscopic Full-Field Diffraction X-ray Microscopy\\}
Stroboscopic full-field diffraction x-ray microscopy experiment was performed at the ID01 beamline of the European synchrotron (ESRF)\cite{Leake2019}. The photon energy was 8 keV. The SAW device was illuminated with a quasi-parallel x-ray beam with a convergent angle of about $10^{-5}$ rad. The pseudo-cubic 300 reflection of \ce{LiNbO3} was chosen with an incident angle of 31.4° and a $2\theta$ angle of 62.8°. The specular reflection allows unambiguous access to the atomic displacements in the direction of the surface normal (\textbf{Y} direction in Figure~\ref{fig:sch}). The diffracted beam was imaged with an objective lens consisting of 50 \ce{Be} parabolical refractive lenses with 50 $\mu$m apex radius of curvature\cite{Snigirev1996}. The sample-to-lens distance was 114 mm. An Andor Zyla 4.2 sCMOS camera with 6.5 $\mu$m pixel size was placed at 6.4 m downstream the objective lens. The effective pixel size was 114 nm in the (v)ertical direction and 202 nm in the (h)orizontal direction. The elongation in the horizontal direction was caused by image projection along the exit wave direction. With each acquisition, a dark-field image of the sample is obtained, corresponding to a field of view of 450(h)$\times$250(v) $\mu$m$^2$. For details about the imaging mechanism of FFDXM, the reader is referred to \cite{Simons2015, Zhou2018}

In the 4-bunch operation mode, four evenly spaced electron bunches circulate in the storage ring at near the speed of light. A delay generator (BCDU8) converts the 352 MHz radial frequency signal from the storage ring to a synchronized 1.42 MHz signal used to trigger the acoustic devices. 1.42 MHz is known as the bunch clock frequency as it corresponds to the frequency of the electron bunches generating the x-ray pulses. Stroboscopic imaging is accomplished by adjusting the electronic delay added to the trigger signal, which can move in fine steps of 11 ps. For an exposure time of 1 sec, for example, the detector accumulates diffraction signals from more than a million x-ray pulses. But because of the synchronization, all these x-ray pulses would probe the exact same snapshot of the device in time. The time resolution of the technique is ultimately limited by the x-ray pulse width at about 100 ps. The excitation signal applied to the SAW devices was generated using a Keysight 81160A signal generator.

At a given delay, $\theta$-$2\theta$ scan was performed by simultaneously moving the sample incident angle by $\Delta \theta$, and both the objective lens and detector by 2$\Delta \theta$. For all the $\theta$-$2\theta$ scans described in this work, the step size was $\Delta \theta =$ 0.001°. A dark field image of the sample was acquired for each acquisition with a typical exposure time of 2.5 s. 

The SAW device is designed to generate Rayleigh waves, which are evanescent waves propagating primarily near the surface\cite{Rayleigh1885}. The amplitude of the SAW decreases exponentially as a function of depth, confining the elastic energy approximately within a depth of one SAW wavelength \cite{Poplavko2019}. The atomic displacement in the \textbf{Y} direction at a depth of 10 $\mu $m is about 1/5 of the atomic displacement at the surface level\cite{Chai2022}. The depth sensitivity of s-FFDXM probe also follows an exponential decay, owing to the absorption of x-rays entering and exiting the crystal. The conditions in our setup were tuned such that the absorption corrected x-ray intensity per unit volume decreased to 1/5 for a depth of 10 $\mu$m leading to a coincidence of the probed depth and the active depth of the SAW. 

\paragraph*{Integrated intensity as an indicator for the SAW amplitude\\}
The presence of SAW induces local bending at the surface. For a crystal with a narrow x-ray rocking curve such as \ce{LiNbO3}, bending leads to an increase in the diffracted intensity due to the broadening of the angular acceptance of the reflection\cite{Ferrari2012}. For weak bending, the integrated intensity $I$ is inversely proportional to the radius of curvature $R$\cite{Ferrari2013, Erola1990}. The equation is simplified in the current case of co-planar Bragg diffraction,
\begin{equation}
I \propto \frac{1}{R} \exp(\frac{-2\mu t}{\sin\theta})
\label{eq_int1}
\end{equation}
where $\mu$ is the linear attenuation coefficient, $t$ is the thickness of the bent crystal, and $\theta$ is the incident angle. For a SAW with a period of $\lambda_\text{SAW}$, traveling in the \textbf{Z} direction (see Figure~S\ref{fig:sup_bend}), the local radius of curvature is inversely proportional to the maximum out-of-plane displacement $\Delta y$, following 
\begin{equation}
\frac{1}{R} \approx \frac{32\Delta y}{\lambda_\text{SAW}^2}
\label{eq_int2}
\end{equation}
Because $\Delta y$ itself is proportional to the maximum strain amplitude $\Delta a/a$, we have thus 
\begin{equation}
I \propto \frac{\Delta a}{a}
\label{eq_int3}
\end{equation}
We shall also evaluate the validity of the weak bending assumption in our case. For a maximum out-of-plane displacement $\Delta y$ of 10 pm, every 5 $\mu$m length of surface is bent at a radius of curvature of $R = 3$ mm . This results in a deformation parameter $\beta$ of about 0.5 $\mu$m$^{-1}$ with 
\begin{equation}
\beta = \frac{2}{R\delta}
\label{eq_int4}
\end{equation}
Here, the Darwin width $\delta$ equals 14 $\mu$rad. The extinction length $\Lambda$ is 4 $\mu$m. The assumption of weak bending is thus valid in our case with $\beta\Lambda\sim1$. We note that even in the strong bending case ($\beta\Lambda>>1$), the diffraction intensity still remains as a good indicator for the SAW amplitude albeit at an non-linear dependence \cite{Malgrange2002}.

\paragraph*{Wave model for the SAW device\\}
The SAW device is designed to generate a standing wave in the resonator area, achieving by reflecting and adding together two waves propagating respectively in the \textbf{--Z} and \textbf{+Z} direction.
\begin{equation}
\psi_{z-}(z, t) = A_{z-} \sin(k_z(z-z_0)+\omega(t-t_0))
\label{eq1}
\end{equation}
\begin{equation}
\psi_{z+}(z, t) = A_{z+} \sin(k_z(z-z_0)-\omega(t-t_0))
\label{eq2}
\end{equation}
Here, $A_{z-}$ and $A_{z+}$ are the amplitudes of the two waves. $k_z$ is the spatial frequency and $z_0$ is the spatial phase. The expected value for $k_z$ is $2\pi/\lambda_\text{SAW}$, where $\lambda_\text{SAW}$ is the acoustic wavelength. $\omega$ is the temporal (angular) frequency and $t_0$ is the temporal phase. The expected value for $\omega$ is $2\pi f_\text{SAW}$ where $f_\text{SAW}$ is the frequency of the burst excitation.

Fitting $\psi_{z-}+\psi_{z+}$ to some parts of the experimental data results in a residual term that is best described as 
\begin{equation}
\psi_t(t) = A_t \cos(2 \omega(t-t_0))
\label{eq3}
\end{equation}
$A_t$ is the amplitude. The residual term does not oscillate along the \textbf{Z} direction, which explains the absence of the spatial frequency parameter in the equation. It is also in phase temporally with the two propagating wave, as it can be fitted with the same parameters $\omega$ and $t_0$. 

The parasitic wave is also a standing wave, formed by adding together two waves propagating in opposite directions (\textbf{--X} and \textbf{+X})
\begin{equation}
\psi_{x-}(x, t) = A_{x-} \sin(k_x(x-x_0)+\omega(t-t_1))
\label{eq4}
\end{equation}
\begin{equation}
\psi_{x+}(x, t) = A_{x+} \sin(k_x(x-x_0)-\omega(t-t_1))
\label{eq5}
\end{equation}
Here, $A_{x-}$ and $A_{x+}$ are the amplitudes of the two waves. $k_x$ is the spatial frequency specific for the parasitic waves and $x_0$ is the spatial phase. The expected value for $k_x$ is $2\pi/\lambda_\text{Par}$, where $\lambda_\text{Par}$ is the period of the parasitic wave at about 11 $\mu$m. $\omega$ is the time (angular) frequency and $t_1$ is the temporal phase. $\omega$ is the same as for $\psi_{x-}$, $\psi_{x+}$, $\psi_{z-}$, $\psi_{z+}$ and $\psi_t$, which confirms the source generator being the one single origin for all the waves described in this model.

\paragraph*{Procedures for wave decomposition analysis\\}
A 4D dataset was acquired at each frequency. The four dimensions are respectively the $Z$, $X$ positions on the sample, the $\theta$-$2\theta$ scan, and the delay time. For a given delay, the 31 dark field images of the same $\theta$-$2\theta$ scan was first aligned using a shift correction algorithm. This corrects for the shift of the sample in the FoV, primarily in the \textbf{X} direction, due to an imperfect alignment of the center of rotation axis. Next, a single strain value was calculated for each aligned sample position, and for each delay value, based on the center of mass 2$\theta$ position using the weighted sum technique. This step reduces the 4D data into three dimensions. 

The spatially resolved and time dependent strain variation was analyzed using the wave model described in the previous section. For the analysis of the primary wave with a modulation along the \textbf{Z} direction, the data was first convoluted in the \textbf{X} direction over one period of the parasitic wave. Least square fitting of the parameters was performed simultaneously on a series of sample points within a sliding window in the \textbf{Z} direction. The size of the sliding window was arbitrary, but was never smaller than 88 pixels which is the size of one acoustic wavelength (10 $\mu$m). The fitting was performed in an iterative manner. The $z_0$ and $t_0$ parameters were first fitted, then again together with $k$ and $\omega$. The parameters $A_0$, $A_1$ and $A_2$ were the last to be fitted. And the fitting starts again with $z_0$ and $t_0$ until a good agreement was reached. The fitting window then slides by 1 pixel in the \textbf{Z} direction to initiate fitting on the next sample location. For the analysis of the parasitic wave with a modulation along the \textbf{X} direction, the data was first convoluted in the \textbf{Z} direction over one period of the primary wave. The fitting procedure is otherwise very similar to what was described for the primary excitation. 

\paragraph*{Strain sensitivity of s-FFDXM\\}
X-ray diffraction typically has a strain resolution of $10^{-5}$, as limited by many factors including the energy bandwidth, beam convergence, motor resolution, detector pixel size and detector distance. The strain resolution is generally thought to be worse when combining an objective lens with a pre-focused incident beam, because x-rays originated from the same sample position but corresponding to slightly different diffracted angles are collected by the lens to form a single pixel on the dark field images. The result is a broadening of the diffraction peak which can be observed on the $\theta$-$2\theta$ scan profile shown in Figure~S\ref{fig:sup_radial}. This seems to be in contradiction with the strain resolution reported in this work. For instance, a strain amplitude on the order of $10^{-6}$ was reported in Figure~S\ref{fig:sup_333h}. Below we discuss some of the main factors that have contributed to the high strain sensitivity of this work and attempt to evaluate the uncertainty of the determined strain value.

As stated before, the penetration depth of the s-FFDXM probe matches the penetration depth of the evanescent SAW. Consequently, only diffraction
from the acoustically strained portion of the \ce{LiNbO3} substrate was measured. In the absence of the diffraction peak from the unstrained portion of the substrate, each $\theta$-$2\theta$ scan profile in Figure~S\ref{fig:sup_radial} contains one single peak, which can then be fitted to a Gaussian line shape (Figure~S\ref{fig:sup_strain}a). Performing Gaussian fitting independently for 4 million detector pixels and at 11 different time delays can be computationally expensive. Instead, the $2\theta$ positions in this work were determined using the center of mass calculations. Figure~S\ref{fig:sup_strain}b shows the $2\theta$ positions determined with least square Gaussian fitting and with center of mass calculations, for 880 pixels spanning 100 $\mu$m in the \textbf{Z} direction at a fixed time delay. The calculated values are in good agreement with each other. Because of the good agreement, we can substitute the uncertainty from the least square fits for the uncertainty from the center of mass calculation, as the latter is not straightforward to evaluate. 

To evaluate the uncertainty from the least square fits, we adopt the boot strap method \cite{Cox1994}. For each $\theta$-$2\theta$ scan profile, 100 sets of synthetic data were generated by adding residuals randomly picked from the least square fits, and then optimized using the same algorithm as used on the actual data. Figure~S\ref{fig:sup_strain}c shows a histogram of the uncertainty for the fit $2\theta$ value, obtained after performing the bootstrap methods on the $\theta$-$2\theta$ scan profiles of the 880 pixels mentioned above. The uncertainty for the fit $2\theta$ value, $\sigma_{2\theta}$, is about $10^{-7}$ rad for a confidence level of 95.44\% at 2$\sigma$. We note that the uncertainty for the determined $2\theta$ shift, $\sigma_{\Delta 2\theta}$ has the same value, as $\sigma_{2\theta} = \sigma_{\Delta 2\theta}$.

The measured strain is directly related to the $2\theta$ shift $\Delta 2\theta$ by 
\begin{equation}
\text{strain} = \frac{\Delta a}{a} = -\frac{\Delta q}{q} = -\frac{\Delta 2\theta}{2 \tan{\frac{2\theta}{2}}} \approx -0.82 \times \Delta 2\theta
\label{eq_2theta1}
\end{equation}
Here, $a$ is lattice parameter of \ce{LiNbO3} along the \textbf{Y} direction. $q$ is the momentum transfer of the 300 reflection. The uncertainty of the determined strain $\sigma_{\text{strain}}$ is thus related to the uncertainty of the $2\theta$ shift $\sigma_{\Delta 2\theta}$ by the same relationship
\begin{equation}
\sigma_{\text{strain}} \approx -0.82 \times \sigma_{\Delta 2\theta} \approx 10^{-7}
\label{eq_2theta2}
\end{equation}

An uncertainty on the order of $10^{-7}$ could explain why leaked SAW with a strain amplitude on the order of $10^{-6}$ could be accurately imaged with s-FFDXM. Interestingly, had the Bragg peak measured with the $\theta$-$2\theta$ scans not been broadened and has, for instance, a FWHM of less than one angular step or 0.002$^\circ$, the Gaussian fitting or the center of mass calculation would have been much less accurate. In a way, the pre-focusing and the use of objective lens make it easier to detect peak shift of sub angular step size, which in turn increased the accuracy of the determined strain amplitude.

\paragraph*{Finite Element Modeling\\}
Finite element modeling of the acoustic mode properties was performed using the COMSOL Multiphysics package. The structure considered is a thin slab corresponding to a portion of one aluminium electrode. The thickness of the slab is $0.05\lambda_\text{SAW}$, where $\lambda_\text{SAW}$ is the periodicity of the electrodes. The slab is located atop a portion of \ce{LiNbO3} substrate, the thickness of which is 5$\lambda_\text{SAW}$. To prevent reflections of waves at the bottom of the modelled volume, a perfectly matched layer is added below the finite \ce{LiNbO3} substrate. A vacuum enclosure is also added over the substrate and around the electrode to provide a realistic electrostatic environment. The crystal orientation of the \ce{LiNbO3} substrate is considered through a rotation of its material frame using Euler angles (0°, 90°, 0°). The periodicity of the electrodes is considered through the application of periodic boundary conditions with a Floquet wave-vector of (0, 0, $2\pi/\lambda_\text{SAW}$) along the edges of the structure perpendicular to the propagation direction (\textbf{Z} direction in Figure~\ref{fig:sch}). The electrodes are considered infinitely long by application of periodic boundary conditions with a Floquet wavevector of (0, 0, 0) along the edges of the structure intersecting the electrode. For Figure~S\ref{fig:sup_pen} and S\ref{fig:sup_disp}, a harmonic calculation was performed, by applying a harmonic voltage of 1 V to the electrode and varying the frequency. The corresponding displacements or strains were extracted at the frequencies of interest after interpolation of the calculated fields on a regular rectangular grid. Otherwise, the electrode was set to 0 V, and an eigenvalue analysis provided the frequencies of the first modes of interest and their modal displacements.

\section*{Acknowledgments}
This work was performed at the ID01 beamline of the European Synchrotron. Work performed at the Center for Nanoscale Materials, a U.S. Department of Energy Office of Science User Facility, was supported by the U.S. DOE, Office of Basic Energy Sciences, under Contract No. DE-AC02-06CH11357. TZ acknowledges Flora Yakhou-Harris and Roberto Homs for help setting up the experiment. TZ acknowledges useful discussion with Michael Wulff and Haidan Wen. AR acknowledges Sylvain Dumas for lending the signal generator during the experiment. AR acknowledges Antoine Nowodzinski, and Patrick Brunet-Manquat for help in setting up the software controls for the signal generator. PGE acknowledges support from the U.S. DOE, Office of Basic Energy Sciences, under contract number DE-FG02-04ER46147. The fabrication of the SAW devices has been performed with the help of Plateforme Technologique Amont de Grenoble, with the financial support of the Nanosciences aux limites de la Nanoélectronique Foundation and the CNRS Renatech network.

\bibliography{Tao}

\begin{thebibliography}{10}

\bibitem{Campbell1989}
C.~Campbell, {\it {Surface Acoustic Wave Devices and Their Signal Processing
  Applications.}\/} (Elsevier Science, 1989), first edn.

\bibitem{Kumar2022}
A.~Kumar, R.~Prajesh, {\it Sensors and Actuators A: Physical\/} {\bf 339},
  113498 (2022).

\bibitem{Lange2008}
K.~L{\"{a}}nge, B.~E. Rapp, M.~Rapp, {\it Analytical and Bioanalytical
  Chemistry\/} {\bf 391}, 1509 (2008).

\bibitem{Yeo2014}
L.~Y. Yeo, J.~R. Friend, {\it Annual Review of Fluid Mechanics\/} {\bf 46}, 379
  (2014).

\bibitem{Delsing2019}
P.~Delsing, {\it et~al.\/}, {\it Journal of Physics D: Applied Physics\/} {\bf
  52}, 353001 (2019).

\bibitem{Dumur2021}
{\'{E}}.~Dumur, {\it et~al.\/}, {\it npj Quantum Information 2021\/} {\bf 7}, 1
  (2021).

\bibitem{Andersson2022}
G.~Andersson, {\it et~al.\/}, {\it PRX Quantum\/} {\bf 3}, 010312 (2022).

\bibitem{White1965}
R.~M. White, F.~W. Voltmer, {\it Applied Physics Letters\/} {\bf 7}, 314
  (1965).

\bibitem{Manenti2017}
R.~Manenti, {\it et~al.\/}, {\it Nature Communications 2017\/} {\bf 8}, 1
  (2017).

\bibitem{Manin2018}
J.~Manin, S.~A. Skeen, L.~M. Pickett, {\it Optical Engineering\/} {\bf 57}, 1
  (2018).

\bibitem{Koskela1999}
J.~Koskela, {\it et~al.\/}, {\it Applied Physics Letters\/} {\bf 75}, 2683
  (1999).

\bibitem{Shu2016}
L.~Shu, {\it et~al.\/}, {\it Sensors\/} {\bf 16} (2016).

\bibitem{Kittmann2018}
A.~Kittmann, {\it et~al.\/}, {\it Scientific Reports\/} {\bf 8} (2018).

\bibitem{Knuuttila2000}
J.~V. Knuuttila, M.~M. Salomaa, P.~T. Tikka, {\it Optics Letters, Vol. 25,
  Issue 9, pp. 613-615\/} {\bf 25}, 613 (2000).

\bibitem{Hisatomi2023}
R.~Hisatomi, {\it et~al.\/}, {\it Physical Review B\/} {\bf 107}, 165416
  (2023).

\bibitem{Hesjedal1998}
T.~Hesjedal, E.~Chilla, H.~J. Fr{\"{o}}hlich, {\it Applied Physics Letters\/}
  {\bf 70}, 1372 (1998).

\bibitem{Hanke2023}
M.~Hanke, {\it et~al.\/}, {\it Phys. Rev. Appl.\/} {\bf 19}, 024038 (2023).

\bibitem{Ludvigsen2015}
H.~Ludvigsen, {\it et~al.\/}, {\it Optics Express, Vol. 23, Issue 8, pp.
  9690-9695\/} {\bf 23}, 9690 (2015).

\bibitem{Nicolas2014}
J.-D. Nicolas, {\it et~al.\/}, {\it Journal of Applied Crystallography\/} {\bf
  47}, 1596 (2014).

\bibitem{Reusch2013}
T.~Reusch, {\it et~al.\/}, {\it AIP Advances\/} {\bf 3}, 072127 (2013).

\bibitem{Whiteley2019}
S.~J. Whiteley, F.~J. Heremans, G.~Wolfowicz, D.~D. Awschalom, M.~V. Holt, {\it
  Nature Communications 2019\/} {\bf 10}, 1 (2019).

\bibitem{Lipiainen2015}
L.~Lipi{\"{a}}inen, K.~Kokkonen, M.~Kaivola, {\it Journal of
  Microelectromechanical Systems\/} {\bf 24}, 1642 (2015).

\bibitem{Telschow2003}
K.~L. Telschow, V.~A. Deason, D.~L. Cottle, J.~D. Larson, {\it IEEE
  Transactions on Ultrasonics, Ferroelectrics, and Frequency Control\/} {\bf
  50}, 1279 (2003).

\bibitem{Koskela2000}
J.~Koskela, V.~Plessky, M.~Salomaa, {\it IEEE Transactions on Ultrasonics,
  Ferroelectrics and Frequency Control\/} {\bf 47}, 1550 (2000).

\bibitem{timeres}
N.~Sévelin-Radiguet, {\it et~al.\/}, {\it urn:issn:1600-5775\/} {\bf 29}, 167
  (2022).

\bibitem{Kalman1983}
Z.~H. Kalman, S.~Weissmann, {\it Journal of Applied Crystallography\/} {\bf
  16}, 295 (1983).

\bibitem{Hossain2019}
M.~M. Hossain, {\it Heliyon\/} {\bf 5}, e01436 (2019).

\bibitem{Kushibiki1999}
J.~Kushibiki, I.~Takanaga, M.~Arakawa, T.~Sannomiya, {\it IEEE Transactions on
  Ultrasonics, Ferroelectrics and Frequency Control\/} {\bf 46}, 1315 (1999).

\bibitem{Inoue2007}
S.~Inoue, {\it et~al.\/}, {\it IEEE Transactions on Ultrasonics,
  Ferroelectrics, and Frequency Control\/} {\bf 54}, 1692 (2007).

\bibitem{Holmgren2007}
O.~Holmgren, {\it et~al.\/}, {\it IEEE Transactions on Ultrasonics,
  Ferroelectrics, and Frequency Control\/} {\bf 54}, 861 (2007).

\bibitem{Howell2001}
K.~B. Howell, {\it Principles of Fourier Analysis\/}  (2001).

\bibitem{Kirchhof2023}
J.~N. Kirchhof, K.~I. Bolotin, {\it npj 2D Materials and Applications 2023\/}
  {\bf 7}, 1 (2023).

\bibitem{Rivas2022}
D.~E. Rivas, {\it et~al.\/}, {\it Optica, Vol. 9, Issue 4, pp. 429-430\/} {\bf
  9}, 429 (2022).

\bibitem{Bajt2018}
S.~Bajt, {\it et~al.\/}, {\it Light, science and applications\/} {\bf 7}, 17162
  (2018).

\bibitem{PC1}
E.~Alonso-Redondo, {\it et~al.\/}, {\it Nature Communications 2015\/} {\bf 6},
  1 (2015).

\bibitem{Bartasyte2017}
A.~Bartasyte, S.~Margueron, T.~Baron, S.~Oliveri, P.~Boulet, {\it Advanced
  Materials Interfaces\/} {\bf 4}, 1600998 (2017).

\bibitem{Kuppel2002}
C.~C. Kuppel, L.~Reindl, R.~Weigel, {\it IEEE Microwave Magazine\/} {\bf 3}, 65
  (2002).

\bibitem{Leake2019}
S.~J. Leake, {\it et~al.\/}, {\it Journal of synchrotron radiation\/} {\bf 26},
  571 (2019).

\bibitem{Snigirev1996}
A.~Snigirev, V.~Kohn, I.~Snigireva, B.~Lengeler, {\it Nature 1996 384:6604\/}
  {\bf 384}, 49 (1996).

\bibitem{Simons2015}
H.~Simons, {\it et~al.\/}, {\it Nature communications\/} {\bf 6}, 6098 (2015).

\bibitem{Zhou2018}
T.~Zhou, {\it et~al.\/}, {\it Microscopy and Microanalysis\/} {\bf 24}, 128
  (2018).

\bibitem{Rayleigh1885}
L.~Rayleigh, {\it Proceedings of the London Mathematical Society\/} {\bf
  s1-17}, 4 (1885).

\bibitem{Poplavko2019}
Y.~M. Poplavko, {\it Electronic Materials\/} pp. 71--93 (2019).

\bibitem{Chai2022}
H.~Y. Chai, E.~J. Chen, Y.~F. Chai, W.~H. Ke, H.~X. Zhu, {\it Soil Dynamics and
  Earthquake Engineering\/} {\bf 156}, 107215 (2022).

\bibitem{Ferrari2012}
C.~Ferrari, {\it Optical Engineering\/} {\bf 51}, 046502 (2012).

\bibitem{Ferrari2013}
C.~Ferrari, E.~Buffagni, E.~Bonnini, D.~Korytar, {\it Journal of Applied
  Crystallography\/} {\bf 46}, 1576 (2013).

\bibitem{Erola1990}
E.~Erola, V.~Eteläniemi, P.~Suortti, P.~Pattison, W.~Thomlinson, {\it Journal
  of Applied Crystallography\/} {\bf 23}, 35 (1990).

\bibitem{Malgrange2002}
C.~Malgrange, {\it Cryst. Res. Technol\/} {\bf 37}, 654 (2002).

\bibitem{Cox1994}
D.~Cox, D.~Hinkley, N.~Reid, D.~Rubin, B.~Silverman, {\it An Introduction to
  the Bootstrap\/} {\bf 21}, 32 (1994).

\end{thebibliography}

\bibliographystyle{Science}
\newpage

\setcounter{figure}{0}
\captionsetup[figure]{
  labelfont=bf,
  labelformat=Sfignum,
  labelsep=colon,
  name=Figure
}

\begin{figure}[!ht]
\centering
\includegraphics[width=0.5\textwidth]{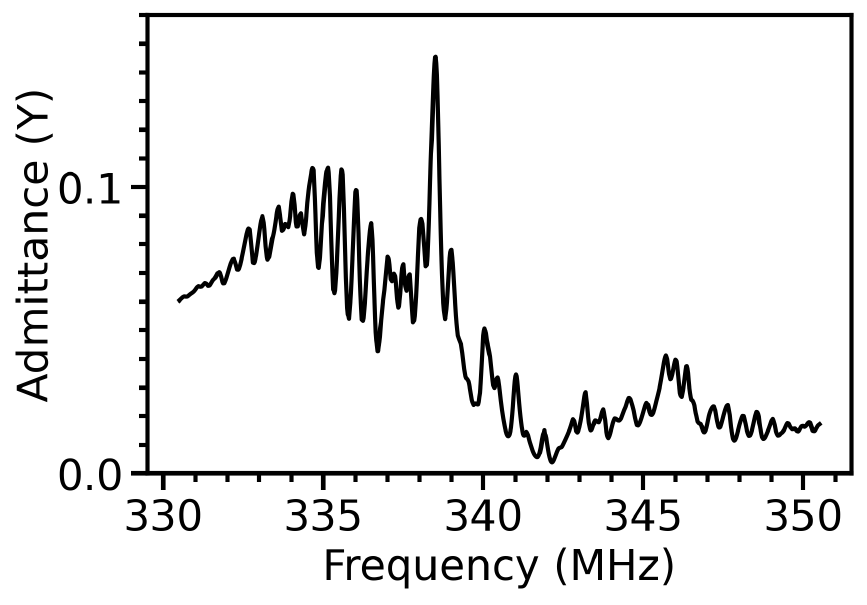}
\caption{\label{fig:sup_imp} \textbf{Electrical verification of the SAW resonance.} The admittance \textit{Y} parameter measured using an Agilent N5230A vector network analyzer, prior to mounting the device for the s-FFDXM experiment. The result shows an anti-resonance frequency at around 342 MHz, a resonance frequency at around 339 MHz.}
\end{figure}
\newpage

\begin{figure}[!ht]
\centering
\includegraphics[width=0.8\textwidth]{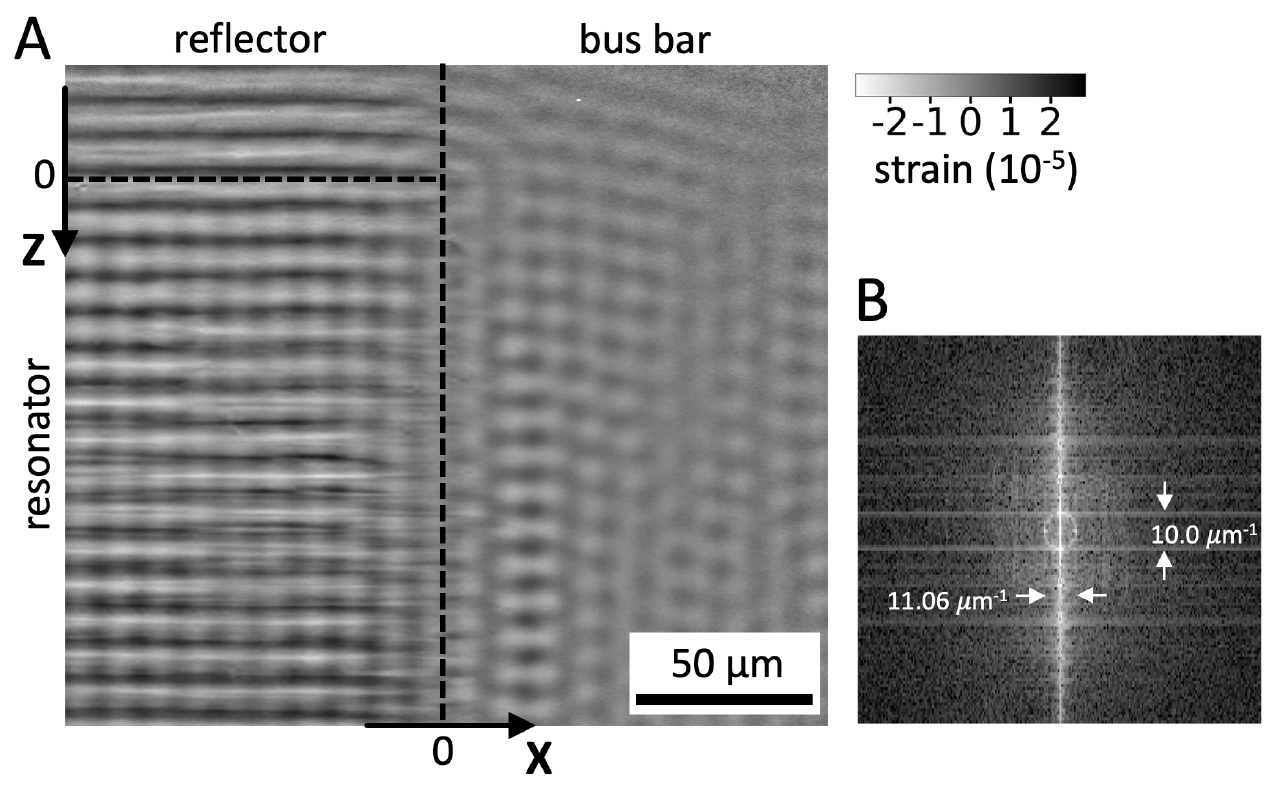}
\caption{\label{fig:sup_fft} \textbf{Evidence of the two-dimensional wave pattern.} (A) Strain map of the SAW device at $t = 143.1$ ns for an excitation frequency of 333 MHz. (B) FFT result on the resonator area for the strain map shown in (A).}
\end{figure}
\newpage

\begin{figure}[!ht]
\centering
\includegraphics[width=0.5\textwidth]{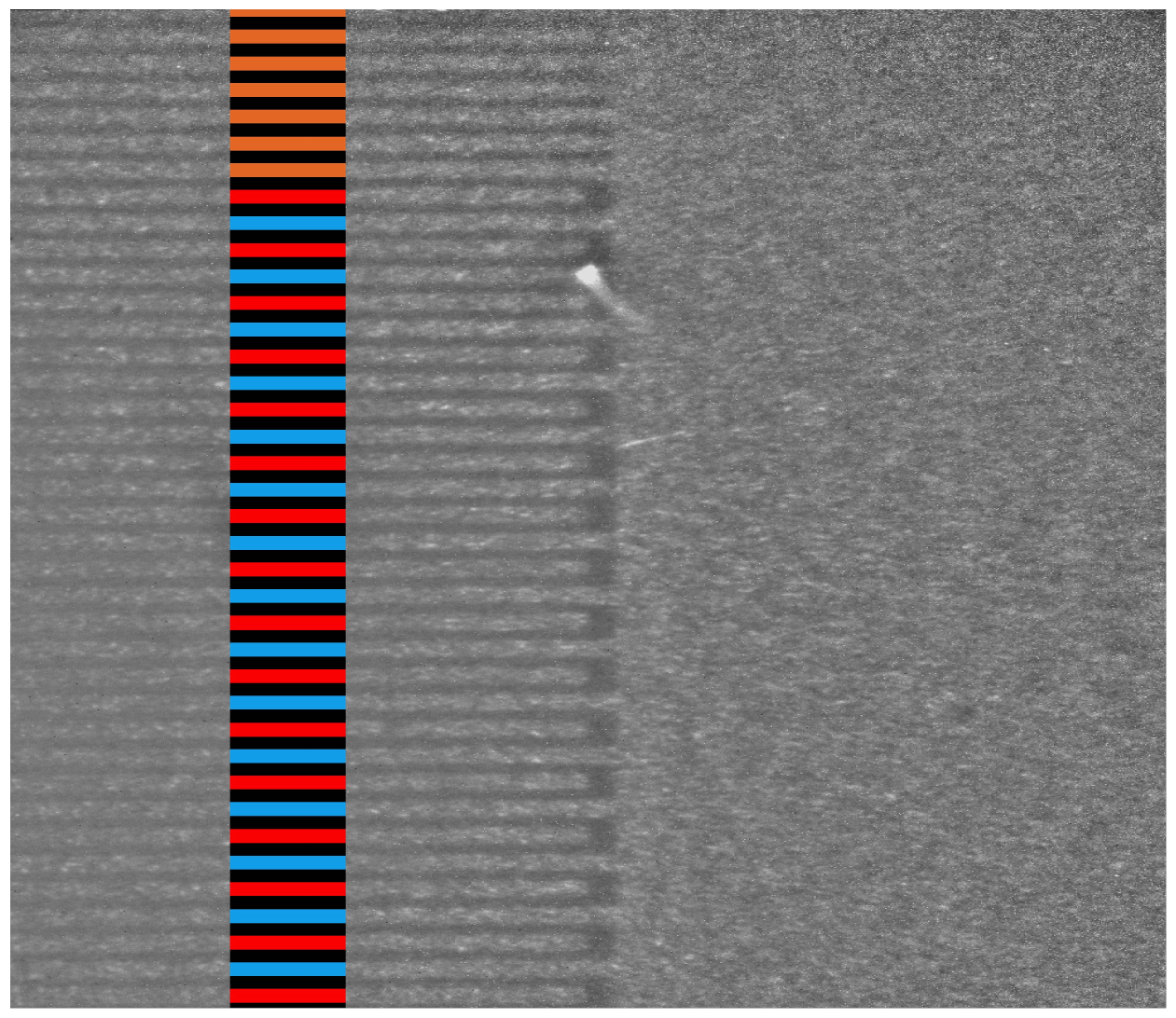}
\caption{\label{fig:sup_ecd} \textbf{Position of the IDT electrodes and reflector gratings.} FFDXM image taken without electrical excitation at $\Delta \theta = -0.2^{\circ}$ off the Bragg peak. At this angle, the metallization layers are visible as they introduces weak contact strain on the \ce{LiNbO3} substrate. The $Z$ positions of the electrode are then extracted and used to identify the reflector area, as well as the nodes and anti-nodes in the resonator area.}
\end{figure}
\newpage

\begin{figure}[!ht]
\centering
\includegraphics[width=0.5\textwidth]{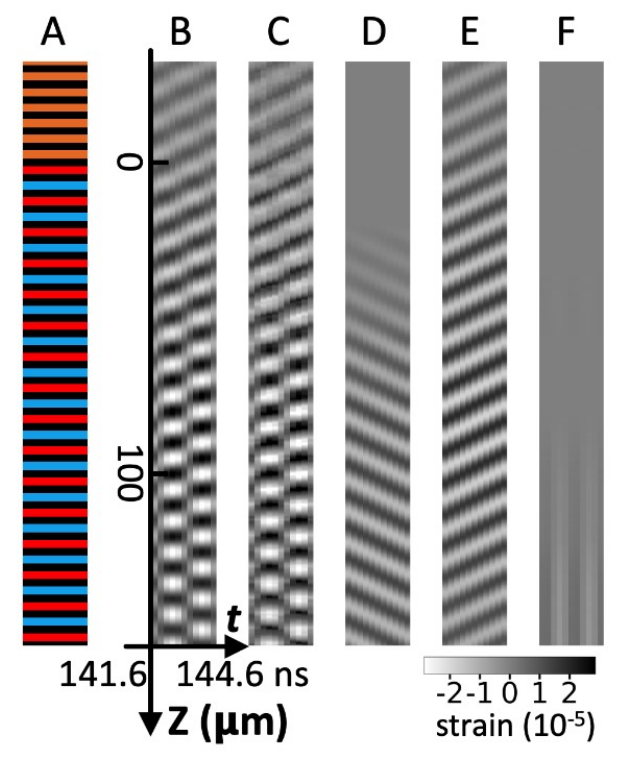}
\caption{\label{fig:sup_333v} \textbf{Spatiotemporal strain map of the primary excitation.} (A) Positions of the reflector gratings (orange) and IDT electrodes (blue and red). (B) Summed contributions from $\psi_{z+}$, $\psi_{z-}$ and $\psi_t$ in the region near $X=-100~\mu$m from $t =$ 141.6 to 144.6 ns. (C) Experimental spatiotemporal strain map previously shown in Figure~\ref{fig:333}B. Spatiotemporal strain map for individual wave components is shown for (D) $\psi_{z+}$ (E) $\psi_{z-}$ and (F) $\psi_t$.}
\end{figure}
\newpage

\begin{figure}[!ht]
\centering
\includegraphics[width=1\textwidth]{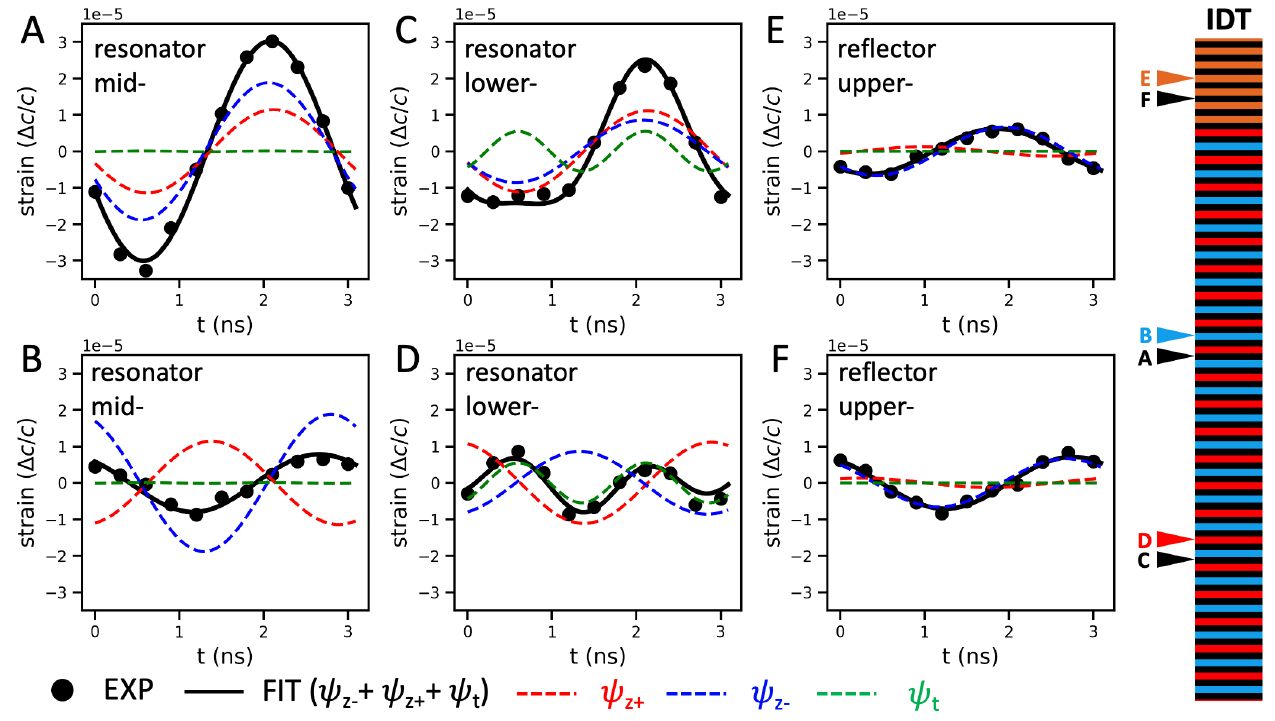}
\caption{\label{fig:sup_fit} \textbf{Selected results from the wave decomposition analysis.} (A) shows the time-dependent strain evolution for an anti-node in the middle of the resonator. The experimental data can be perfectly described by two propagating waves $\psi_{z-}$ and $\psi_{z+}$ in phase with each other. (B) shows the time-dependent strain evolution for a node in the middle of the resonator. The experimental data can be perfectly described by two propagating waves completely out-of-phase with each other. (C) shows the time-dependent strain evolution for an anti-node at $Z>100~\mu$m. The experimental data no longer matches the simple summation of two propagating waves. The time dependent strain evolution requires a third term $\psi_{t}$. (D) The existence of this third term is perhaps more obvious for the experimental data on a nearby node. Here, the two out-of-phase waves cancel each other, leaving behind a wave that can only be fitted with a term with a doubled temporal frequency. The wave becomes less stationary as one gets close to the reflector area. Inside the reflector area, the amplitude of $\psi_{z+}$ becomes negligible and the experimental data can be described entirely by the leaked wave $\psi_{z-}$ propagating in the \textbf{--Z} direction, whether it is on the reflector grating (E) or in-between them (F).}
\end{figure}
\newpage

\begin{figure}[!ht]
\centering
\includegraphics[width=0.5\textwidth]{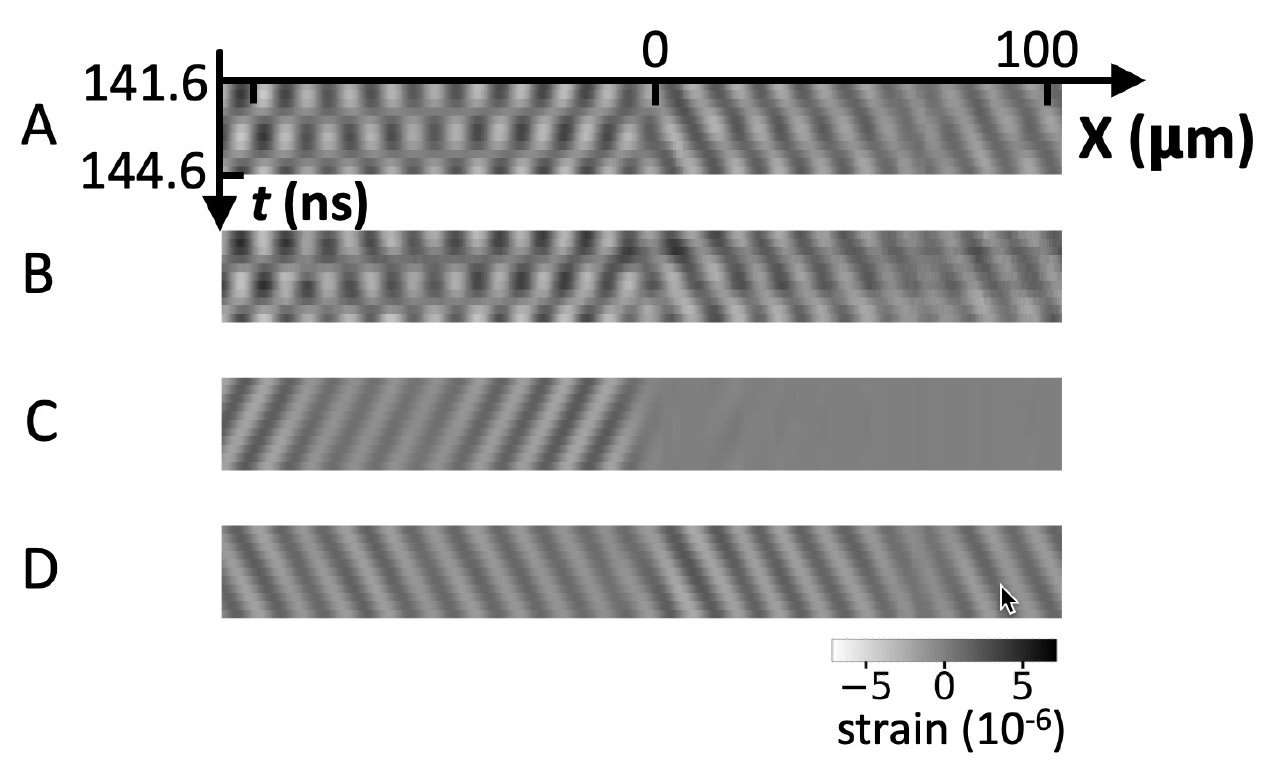}
\caption{\label{fig:sup_333h} \textbf{Spatiotemporal strain map of the parasitic excitation.} (A) Summed contributions from $\psi_{x+}$ and $\psi_{x-}$ in the region near $Z=30~\mu$m from $t =$ 141.6 to 144.6 ns. (B) Experimental spatiotemporal strain map previously shown in Figure~\ref{fig:333}C. The colormap is rescaled to improve the visibility of the weak wave amplitudes. Spatiotemporal strain map for individual wave components is shown for (C) $\psi_{x-}$ and (D) $\psi_{x+}$.}
\end{figure}
\newpage

\begin{figure}[!ht]
\centering
\includegraphics[width=0.5\textwidth]{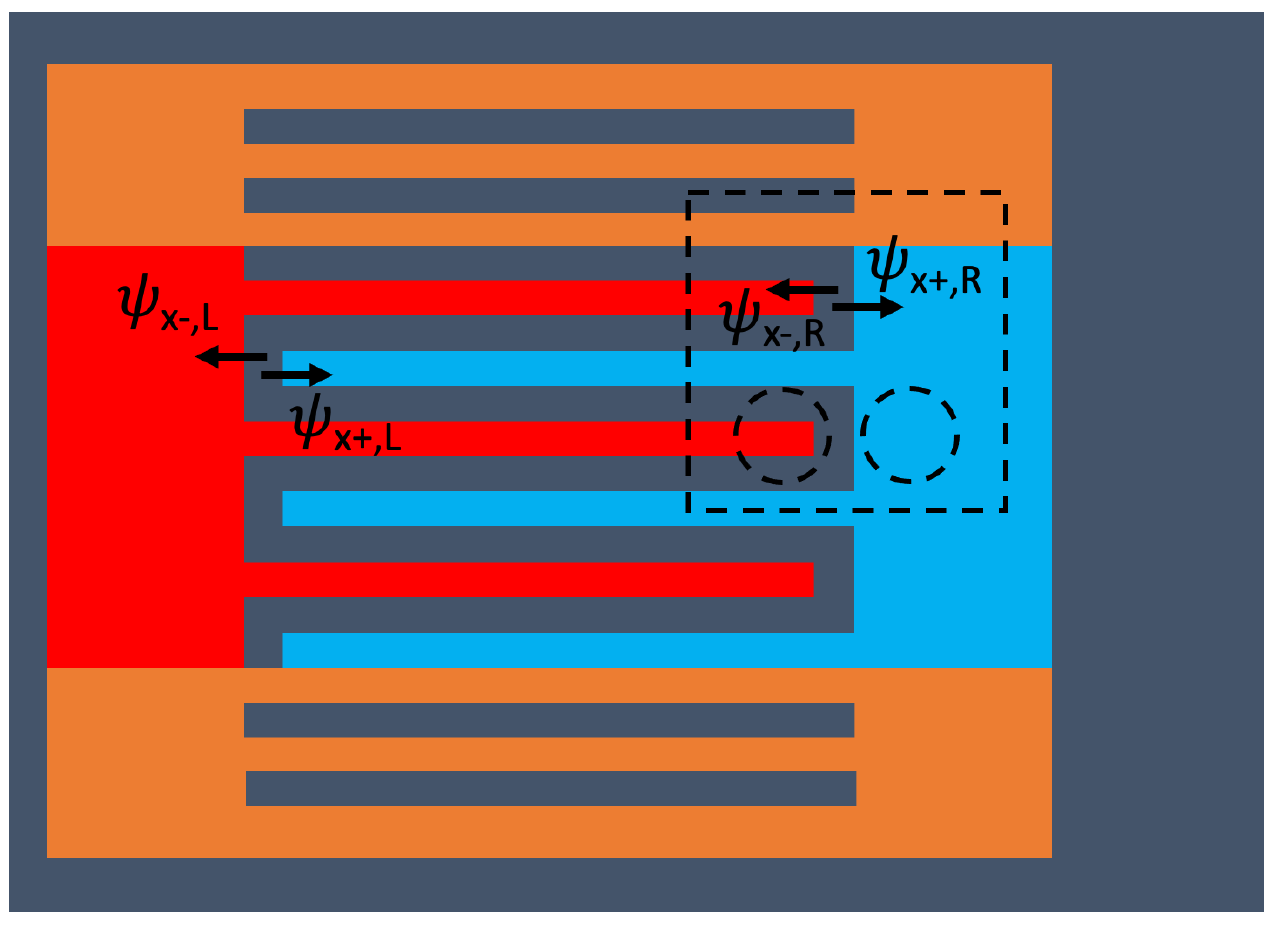}
\caption{\label{fig:sup_ori} \textbf{Origin of the parasitic excitation.} The parasitic excitations were created between the tip of one IDT finger and the opposite busbar, which are marked by the dashed circles. The FoV of the experiment is schematically marked by the dashed rectangle. $\psi_{x-,R}$ was created at the boundary between the resonator and the bus bar on its right, and was responsible for the portion of $\psi_{x-}$ observed in Figure~S\ref{fig:sup_333h}c at $X<0$. $\psi_{x+,R}$ was also created at the boundary between the resonator and the bus bar on its right, and was responsible for the portion of $\psi_{x+}$ observed in Figure~S\ref{fig:sup_333h}d at $X>0$. $\psi_{x+,L}$ was created at the boundary between the resonator and the bus bar on its left. Initially outside of the FoV, $\psi_{x+,L}$ propagated in the \textbf{+X} direction, and was responsible for the portion of $\psi_{x+}$ observed in Figure~S\ref{fig:sup_333h}d at $X<0$.}
\end{figure}
\newpage

\begin{figure}[!ht]
\centering
\includegraphics[width=0.8\textwidth]{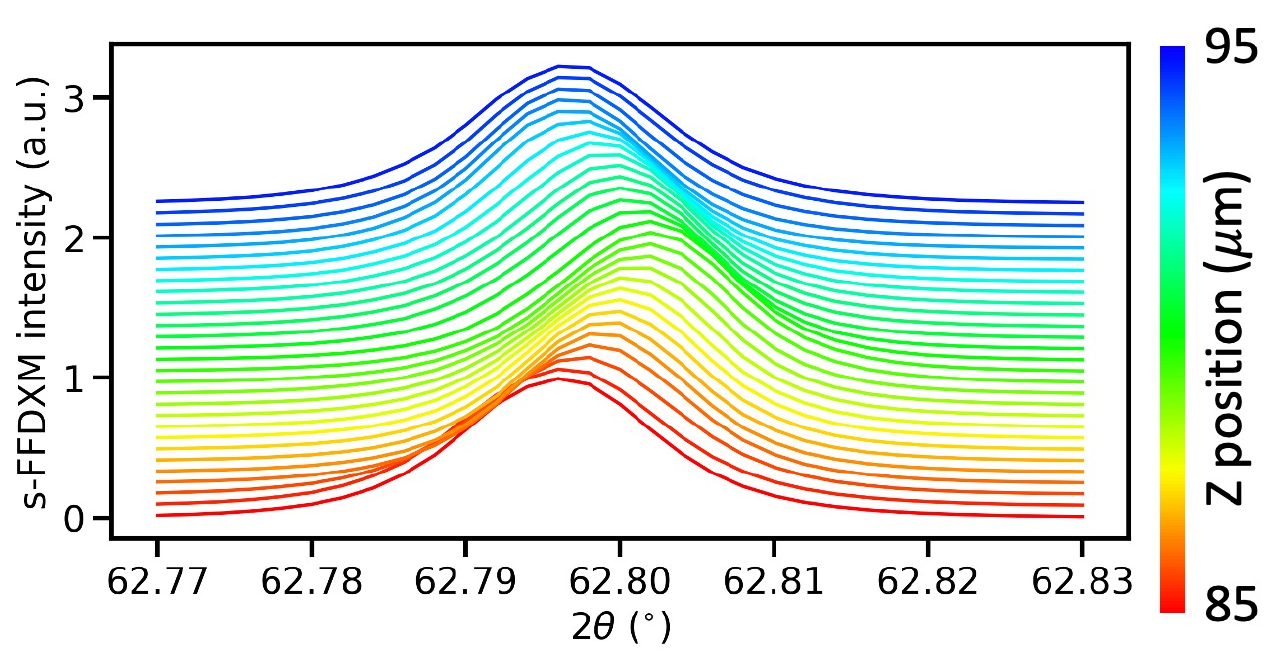}
\caption{\label{fig:sup_radial} \textbf{Data quality of the $\theta$-$2\theta$ scans.} Per pixel $\theta$-$2\theta$ scan curve at $t = 143.1$ ns for an excitation frequency of 333 MHz. The selected pixels covers an entire spatial period of $\lambda_\text{SAW} = 10~\mu$m in the resonator along the \textbf{Z} direction. Compressive (tensile) strain shifts the intensity maximum of the Bragg reflection to a higher (lower) value of $2\theta$. The $\theta$-$2\theta$ scan curves are normalized and shifted vertically for visibility.}
\end{figure}
\newpage

\begin{figure}[!ht]
\centering
\includegraphics[width=0.5\textwidth]{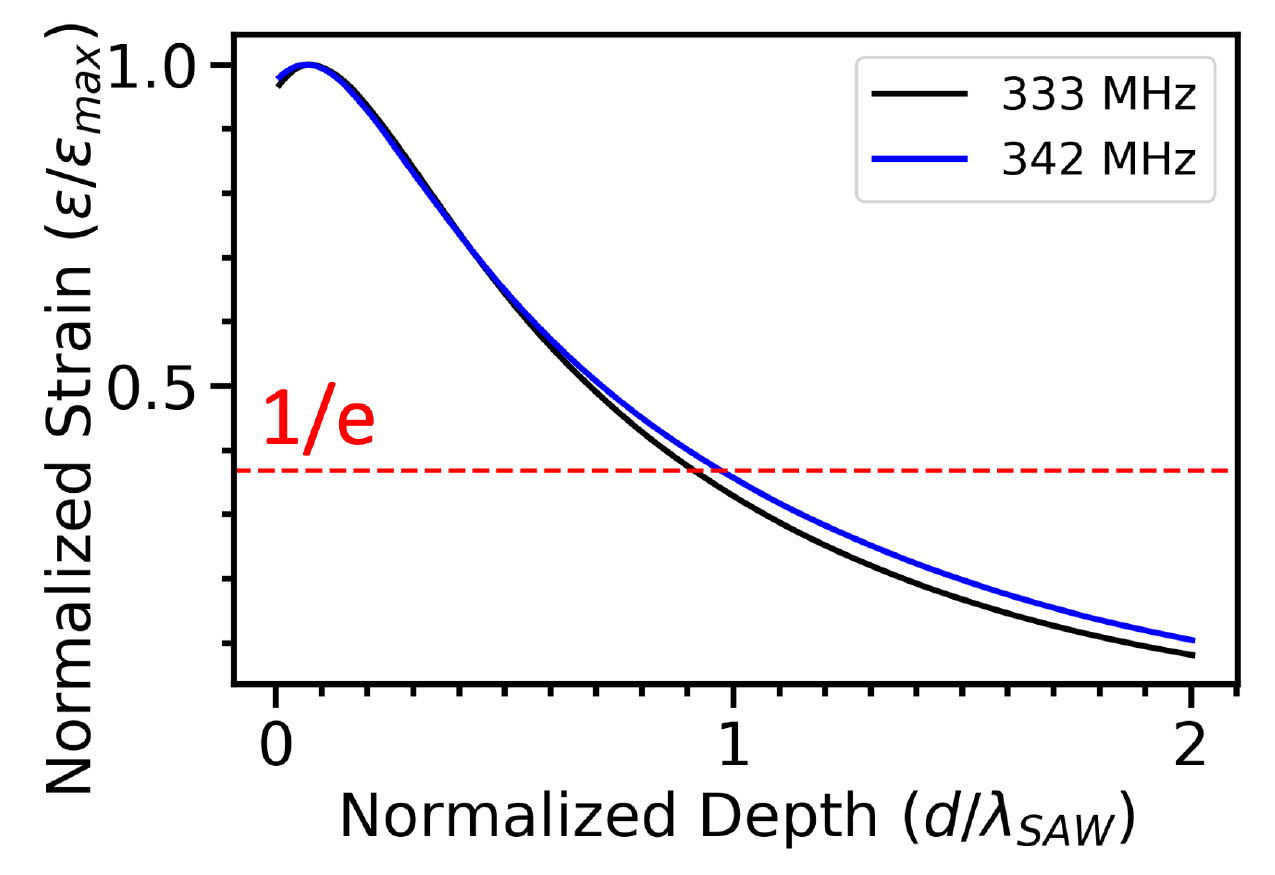}
\caption{\label{fig:sup_pen} \textbf{FEM calculated SAW amplitude as a function of depth.} The SAW amplitude is normalized to its maximum value while the penetration depth is normalized by the device period $\lambda_\text{SAW}$. It can be seen that the SAW amplitude drops to $1/e$ of its maximum value at around the depth of one $\lambda_\text{SAW}$ for both 333 and 342 MHz.}
\end{figure}
\newpage

\begin{figure}[!ht]
\centering
\includegraphics[width=0.5\textwidth]{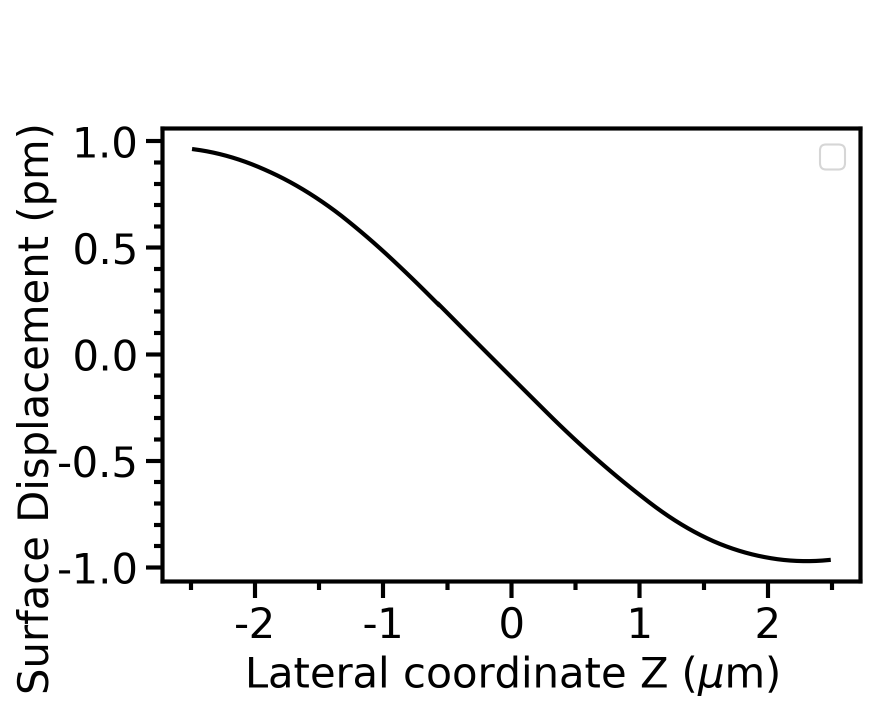}
\caption{\label{fig:sup_disp} \textbf{FEM calculated surface displacement at 333 MHz.} For a strain of $10^{-7}$, the magnitude of the maximum displacement is about 1 pm.}
\end{figure}
\newpage

\begin{figure}[!ht]
\centering
\includegraphics[width=0.5\textwidth]{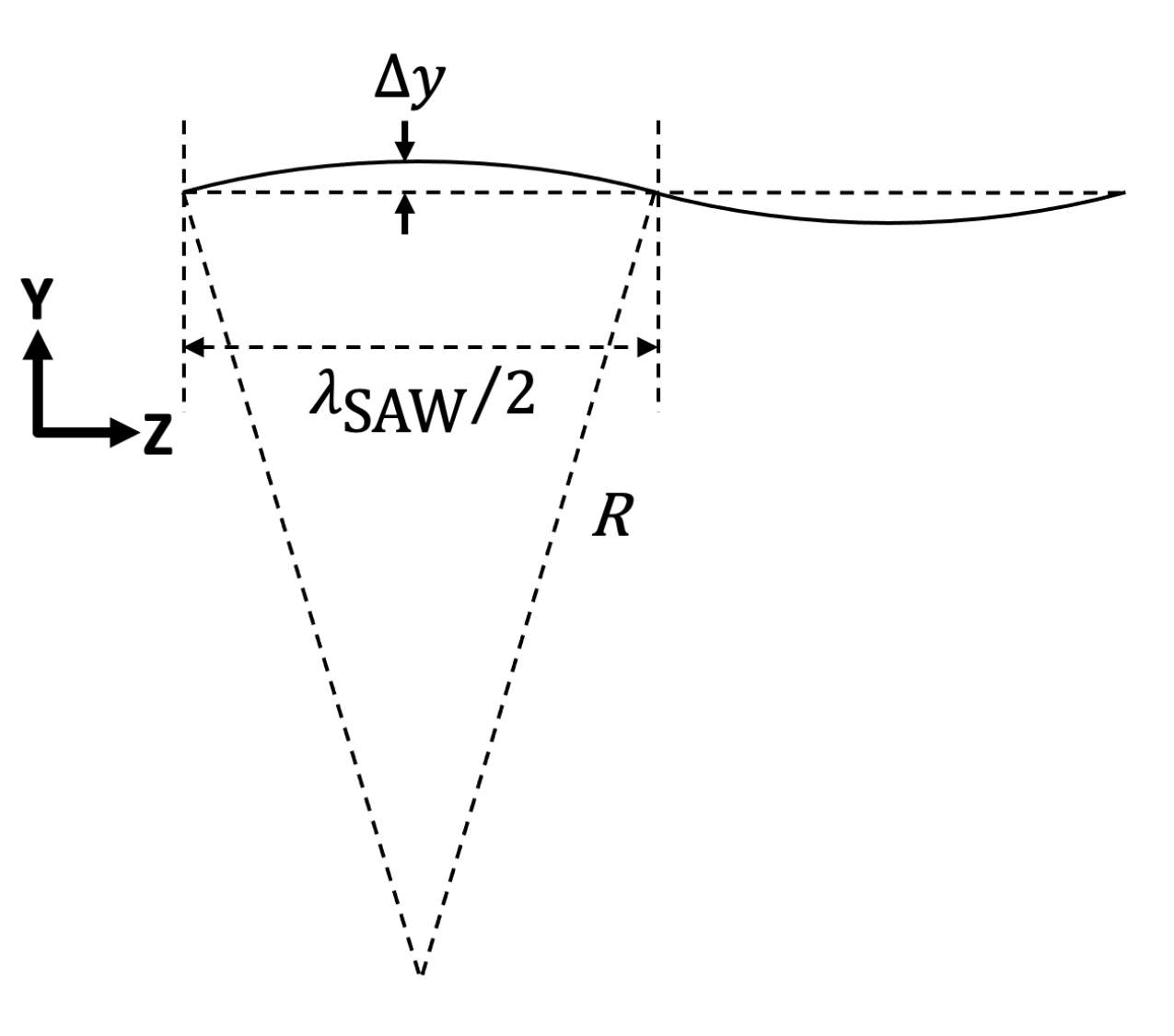}
\caption{\label{fig:sup_bend} \textbf{Schematic for calculating the radius of curvature at the surface.} The SAW travels in the \textbf{Z} direction at a period of $\lambda_\text{SAW}$, causing surface displacement in the \textbf{Y} direction.}
\end{figure}
\newpage

\begin{figure}[!ht]
\centering
\includegraphics[width=0.7\textwidth]{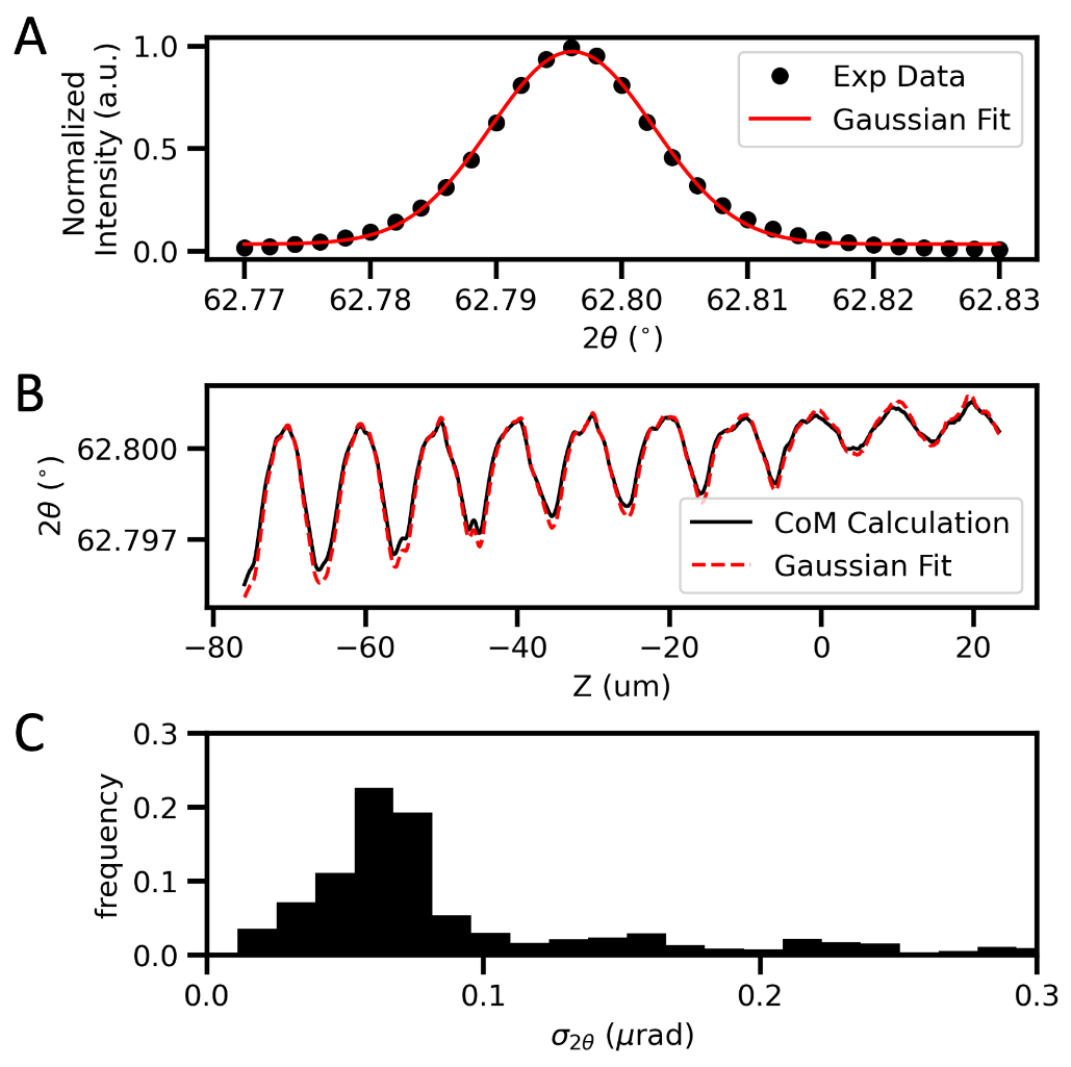}
\caption{\label{fig:sup_strain}  \textbf{Strain sensitivity of s-FFDXM.} (A) Experimental $\theta$-$2\theta$ scan profile versus least square fit result with a Gaussian curve. The experimental data is the same as the first curve shown in Figure~S\ref{fig:sup_radial}. (B) Comparison between the $2\theta$ positions determined with least square Gaussian fitting and with center of mass calculations, for 880 pixels spanning 100 $\mu$m in the \textbf{Z} direction at a fixed time delay. The selected pixels covered both the resonator and the reflector area. (C) Histogram of the uncertainty for the fit $2\theta$ value, obtained after performing the bootstrap methods on the $\theta$-$2\theta$ scan profiles of the 880 pixels mentioned above. The distribution of the bootstrapped parameter was assumed to be normal. The uncertainty was calculated for a confidence level of 95.44\% (i.e., 2$\sigma$).}
\end{figure}
\newpage

\captionsetup[movie]{
  labelfont=bf,
  labelformat=Sfignum,
  labelsep=colon,
  name=Movie
}

\begin{movie}[!ht]
 \fbox{place holder for stack3D\_333.mp4}
\caption{\justifying Strain wave at 333 MHz at the triple boundary separating the resonator, the reflector and the bus bar. A uniform standing wave is observed in the resonator area at $Z>0$ and $X<0$. A weak propagating wave is observed in the reflector area at $Z<0$. Minor leakage is observed in the bus bar area at $X>0$. The video is looped for 5 times.}
\end{movie}

\begin{movie}[!ht]
 \fbox{place holder for stack3D\_342.mp4}
\caption{\justifying Strain wave at 342 MHz at the triple boundary separating the resonator, the reflector and the bus bar. The excited SAW amplitude is not uniform and is strongest near $X=0$. There is no clear difference between the SAW in the resonator area ($Z>0$ and $X<0$) and in the reflector ($Z<0$ and $X<0$). Strong leakage is observed in the bus bar area at $X>0$. The video is looped for 5 times.}
\end{movie}

\begin{movie}[!ht]
 \fbox{place holder for stack\_333.mp4}
\caption{\justifying Strain wave at 333 MHz at the triple boundary separating the resonator, the reflector and the bus bar. The FoV and color scale are the same as in Fig. 3. The video is looped for 5 times.}
\end{movie}

\begin{movie}[!ht]
 \fbox{place holder for stack\_342.mp4}
\caption{\justifying Strain wave at 342 MHz at the triple boundary separating the resonator, the reflector and the bus bar. The FoV and color scale are the same as in Fig. 4. The video is looped for 5 times.}
\end{movie}
\newpage

\end{document}